\documentclass[review,12pt,authoryear]{elsarticle}

\usepackage[latin1]{inputenc}
\usepackage[T1]{fontenc}
\usepackage{textcomp} 
\journal{icarus}
\usepackage{amssymb,amsfonts,amsmath,multirow}
\usepackage{rotating}
\usepackage{graphicx}
\usepackage{framed}
\usepackage{fancybox}
\usepackage{color}
\usepackage{longtable}
\usepackage{subfigure}

\def\arcsec{\hbox{$^{\prime\prime}$}}
\DeclareMathOperator{\Tr}{Tr}

\begin{document}

\begin{frontmatter}



\title{Stratospheric aftermath of the 2010 Storm on Saturn as observed by the TEXES instrument. I. Temperature structure.}


\author[label1,label2]{Thierry Fouchet\corref{cor1}}
\ead{Thierry.Fouchet@obspm.fr}
\author[label3]{Thomas K.\ Greathouse\tnoteref{footnote1}}
\author[label4]{Aymeric Spiga}
\author[label5]{Leigh N.\ Fletcher}
\author[label4]{Sandrine Guerlet}
\author[label6,label7]{Jérémy Leconte}
\author[label8]{Glenn S. Orton}

\address[label1]{LESIA, Observatoire de Paris, PSL Research University, CNRS, Univ. Paris Diderot, Sorbonne Paris Cité, 5 Place Jules Janssen, Meudon, France}
\address[label2]{Sorbonne Universités, UPMC Univ. Paris 06, UMR 8109, LESIA, Meudon, France}
\address[label3]{Southwest Research Institute, Division 15, 6220 Culebra Road, San Antonio, TX 78228, USA}
\address[label4]{Sorbonne Universités, UPMC Paris 06, UMR 8539, LMD, F-75005 Paris, France}
\address[label5]{Department of Physics and Astronomy, University of Leicester, University Road, 
Leicester, LE1 7RH, UK}
\address[label6]{Univ.\ Bordeaux, LAB, UMR 5804, F-33270, Floirac, France}
\address[label7]{CNRS, LAB, UMR 5804, F-33270 Floirac, France}
\address[label8]{Jet Propulsion Laboratory, California Institute of Technology, 4800 Oak Grove Drive, Pasadena, CA, 91109, USA}

\cortext[cor1]{Corresponding author}
\tnotetext[footnote1]{Visiting Astronomer at the Infrared Telescope Facility, which is operated by the University of Hawaii under contract NNH14CK55B with the National Aeronautics and Space Administration.}

\begin{abstract}
We report on spectroscopic observations of Saturn's stratosphere in July 2011 with the Texas Echelon Cross Echelle Spectrograph (TEXES) mounted on the NASA InfraRed Telescope Facility (IRTF). The observations, targeting several lines of the CH$_4$ $\nu_4$ band and the H$_2$ S(1) quadrupolar line, were designed to determine how Saturn's stratospheric thermal structure was disturbed by the 2010 Great White Spot. A study of Cassini Composite Infrared Spectrometer (CIRS) spectra had already shown the presence of a large stratospheric disturbance centered at a pressure of 2~hPa, nicknamed the beacon B0, and a tail of warm air at lower pressures  (Fletcher et al.\ 2012. Icarus 221, 560--586). Our observations confirm that the beacon B0 vertical structure determined by CIRS, with a maximum temperature of $180\pm1$K at 2~hPa, is overlain by a temperature decrease up to the 0.2-hPa pressure level. Our retrieved maximum temperature of $180\pm1$K is colder than that derived by CIRS ($200\pm1$K), a difference that may be quantitatively explained by terrestrial atmospheric smearing. We propose a scenario for the formation of the beacon based on the saturation of gravity waves emitted by the GWS. Our observations also reveal that the tail is a planet-encircling disturbance in Saturn's upper stratosphere, oscillating between 0.2 and 0.02~hPa, showing a distinct wavenumber-2 pattern. We  propose that this pattern in the upper stratosphere is either the signature of thermal tides generated by the presence of the warm beacon in the mid-stratosphere, or the signature of Rossby wave activity.

\end{abstract}


\end{frontmatter}

\section{Introduction}

Since at least 1876, each Saturnian year has witnessed the eruption of a Great White Spot (GWS), an event starting from a small convective plume that rapidly grows in area to eventually encircle a whole latitudinal band with a tail of bright clouds \citep{SanchezLavega1987,SanchezLavega1994}. All the documented events have occurred in Saturn's northern hemisphere. The current Saturnian year is no exception, with an eruption that was first detected from ground-based telescopes \citep{SanchezLavega2011} and from the Cassini spacecraft \citep{Fischer2011} on December 5th, 2010. However, the 2010 GWS exhibited two sharp differences compared to the historical record. First, it is the earliest GWS ever detected in Saturn's seasonal cycle. It started at a solar longitude of $L_s=16$°, whereas the previous events occurred in the $L_s=106$°--$170$° interval \citep{SanchezLavega2012}. Second, for the first time, a large thermal and chemical stratospheric disturbance associated with a GWS was detected from ground-based and Cassini infrared observations \citep{Fletcher2011}.

As reported by \citet{SanchezLavega2011,SanchezLavega2012} and \citet{Sayanagi2013}, the storm started at $37.7\pm$0.8°N in planetographic latitude as a small convective bright cloud that was growing rapidly. Two weeks latter, the convective head had grown to an area of $6\times10^7$~km$^2$, was centered at $41.1\pm1.1$°N, and moved westward 10~$\text{m}\, \text{s}^{-1}$ faster than the background wind. This difference in velocity between the convective head and the background atmosphere created a tail of bright clouds moving with the ambient wind, which surrounded the entire 25°N--45°N latitudinal band in about fifty days.
 The convective nature of the storm head was supported by the independent observations of three Cassini instruments. The Radio and Plasma Wave Science (RPWS) instrument detected numerous lightning discharges, indicative of moist convection, originating from the storm region \citep{Fischer2011,Dyudina2013}.  The Visual and Infrared Mapping Spectrometer (VIMS) detected fresh ammonia ice within the head clouds \citep{Sromovsky2013}. Finally, the Composite InfraRed Spectrometer (CIRS) measured a decrease of the hydrogen para fraction at the latitude of the storm, indicative of an upward transport of hydrogen \citep{Achterberg2014}. The convective activity ceased between the 15th and 19th of June 2011, when the storm head encountered the Dark Spot, a persistent dark anticyclonic vortex embedded within the tail, although resurgences of lightning discharges occurred sporadically up to the end of 2011 \citep{SanchezLavega2012,Sayanagi2013}.

\citet{Fletcher2012} gave a comprehensive overview of the 3D structure of the stratospheric thermal and chemical disturbance and of its temporal evolution. It was first detected on January 19th, 2011, from  VLT/VISIR and Cassini/CIRS observations in thermal hydrocarbons bands at 8.6 and 12.3~\textmu m. The thermal images revealed the presence of two stratospheric warm vortices, also nicknamed beacons, centered at a latitude of about 30°N and extending from 18°N up to 55°N. Subsequent Cassini/CIRS, VLT/VISIR and IRTF/MIRSI observations showed that the two beacons strengthened their anomalously high temperatures while they drifted at different rates with respect to System III longitudes, and eventually met and merged in April 2011. The temperature of the merged beacon increased until May 2011, where a maximum temperature of 221.6$\pm1.4$~K at the 2-hPa level was detected. The temperature then started to decay at a rate of about 0.1$\pm0.05$K per day, but the thermal anomaly was still apparent in March 2012. After merger, the beacon accelerated its westward drift rate from (1.6°$\pm$0.2°)/day to (2.7°$\pm$0.04°)/day in late June or early July 2011. The 2010 GWS did not only affect the stratospheric thermal structure but also perturbed the stratospheric chemistry \citep{Cavalie2015,Moses2015}. Using Cassini/CIRS and McMath-Pierce/Celeste spectra, \citet{Fletcher2012}, \citet{Hesman2012}, and \citet{Moses2015} reported an increase in volume mixing ratio within the beacon for acetylene, ethane and ethylene.

Due to their medium spectral resolution ($R<2,500$), the Cassini/CIRS nadir spectra presented by \citet{Fletcher2012} are limited in sensitivity to the 500--0.5~hPa pressure range. Hence, how the stratospheric thermal structure was affected by the 2010 GWS above the 0.5-hPa pressure level remains unknown. This situation also hampers the accurate measurement of the chemical anomaly associated with the beacon, because the sensitivity range for ethane, ethylene, and acetylene extends to higher altitudes than that probed by methane in CIRS nadir spectra. This lack of information on the full thermal structure of the beacon affects our ability to propose a forcing mechanism at the origin of the stratospheric disturbances and to determine how the upper stratosphere was affected by the tropospheric GWS or by the beacon itself.

Two different techniques could be used to probe Saturn's upper stratosphere. \citet{Guerlet2013} presented CIRS observations obtained in limb-viewing geometry, hence sounding up to the 0.01-hPa pressure level. Here, we present high spectral resolution observations of Saturn obtained with the Texas Echelon Cross Echelle Spectrograph (TEXES) mounted on the NASA InfraRed Telescope Facility (IRTF). \citet{Greathouse2005} demonstrated that TEXES observations of hydrocarbon thermal emissions could be used to probe the temperature and the chemical composition up to the 0.01-hPa pressure level. Section~\ref{SecObs} presents the instrument and the observations. Section~\ref{SecMethods} presents the radiative transfer code and the retrieval methods leading to the results presented in Sec.~\ref{SecResults} and discussed in Sec.~\ref{SecDiscussion}.

 \section{Observations} 
 \label{SecObs}
 
 Taking advantage of Director Discretionary Time received from the NASA Infrared Telescope Facility (IRTF), we obtained high-spectral resolution observations of Saturn and the beacon using TEXES, the Texas Echelon crossed-dispersed Spectrograph \citep{Lacy2002}, between July 15th and July 20th, 2011. At this period, Saturn's angular diameter was about 17\arcsec. The observations were primarily taken during afternoon hours between 2 pm and 7 pm Hawaii Standard Time (HST), however on the night of the 15$^{\text{th}}$ we were allowed to observe until 9 pm HST. Over the 6 days, we observed a number of wavelengths to infer stratospheric and tropospheric temperatures, mixing ratios of key hydrocarbons, and to look for previously undetected hydrocarbons. In this paper, we focus on the temperature retrievals and what they tell us about the vertical structure of the beacon, leaving the hydrocarbon maps and searches for trace constituents to a second follow-on paper. Four spectral settings were used to probe Saturn's temperature, the three first settings at 1230 cm$^{-1}$, 1245 cm$^{-1}$, and 1280 cm$^{-1}$ capturing several strong and weak CH$_4$ emission features sensitive to the stratospheric temperature structure between 10 and 0.01 hPa, and the fourth setting at 587 cm$^{-1}$ capturing both the strong H$_2$ S(1) quadrupolar emission as a probe of the stratospheric temperature near 2 hPa, and the collision-induced continuum, which probes the troposphere at 100 hPa (see Table~\ref{TabObs} for a summary). These CH$_4$ settings used the 1.4\arcsec$\times$6\arcsec\ slit, and the H$_2$ setting used the 2\arcsec$\times$11\arcsec\ slit. The wider slit is used at 587 cm$^{-1}$ due to diffraction making the telescope's FWHM larger at longer wavelengths. Under good observing conditions, the slit width would have defined the angular resolution of our observations. However, since we operated during afternoon hours, our effective angular resolution was decreased by the seeing, which was not monitored simultaneously. Hence, we cannot expect a spatial resolution better than 12° of latitude/longitude on the central meridian and at the latitude of the beacon for the CH$_4$ setting, and 17° for the H$_2$ setting.
 
 \begin{table}
 \begin{tabular}{lcccr}
 Observing Night & Sub Earth & Beacon   & Coverage  & Spectral\\ 
 July 2011 (UT) & longitude & longitude at  &  (cm$^{-1}$) & Resolving  \\ 
& in degrees & 0 UT on date &  &  power \\ \hline
 15$^{\text{th}}$ 2:37--3:40 & 351°--27° & 46.3° & 1244.6--1250.5 & 75,000 \\
 15$^{\text{th}}$ 4:41--5:04  & 61°--74° & 46.3° & 586.0--589.7 & 50,000 \\
 17$^{\text{th}}$ 3:16--4:02 & 194°--220° & 51.7° & 1245.4--1250.5 & 75,000 \\ 
 18$^{\text{th}}$ 3:21--4:56 & 288°--341° & 54.4°  &1228.7--1233.9 & 75,000 \\
 19$^{\text{th}}$ 0:46--1:35 & 291°-319° & 57.1° & 585.4--588.4 & 50,000 \\
 19$^{\text{th}}$ 3:58--4:57 & 39°-73° & 57.1° &1276.5--1283.8 &75,000 \\
 20$^{\text{th}}$ 3:57--4:52 & 129°--160° & 59.8° & 1242.8--1249.4 & 75,000  \\ \hline    
 \end{tabular}
 \caption{List and characteristics (date, sub-Earth point and Beacon longitudes, spectral coverage, and spectral resolution) of the TEXES/IRTF data analyzed in this study.\label{TabObs}}
 \end{table}
 
All the observations were retrieved in high-resolution scan mode with the long axis of the slit oriented along celestial north south. Each scan of Saturn started with the slit centered on the beacon latitude at Saturn's central meridian (CM). Due to the rings of Saturn, we employed a special scan pattern on the sky where we offset from the starting position by 30\arcsec\ north. We then sat and took 5 sky frames for use as sky subtraction for the rest of the scan. After taking the sky frames, we offset the slit 12\arcsec\ east and 30\arcsec\ south of the sky position, placing us back in line with the beacon in the North/South direction, but offset from the CM of Saturn by 12\arcsec\ east. We then stepped  the slit west in 0.7\arcsec\ increments across Saturn, Nyquist sampling the width of the slit, until we were 12\arcsec\ west of the CM. The slit would then return to the starting point. On average, we took a set of flat/calibration observations for every 4 scans. The flat field is retrieved while at the initial sky position of the scan, and consists of an observation of a room temperature blackbody, which is positioned just outside the Dewar window, followed by an observation of the night sky. We then use this information to calibrate the data and perform a first order sky correction \citep{Lacy2002}. TEXES absolute calibration is estimated to lie within $\pm20\%$ from comparisons of observed fluxes of standard stars measured over many observing runs.

\begin{figure}[ht!]
\subfigure{ 
\includegraphics[width=0.45\linewidth, clip=true, trim=23mm 65mm 23mm 65mm]{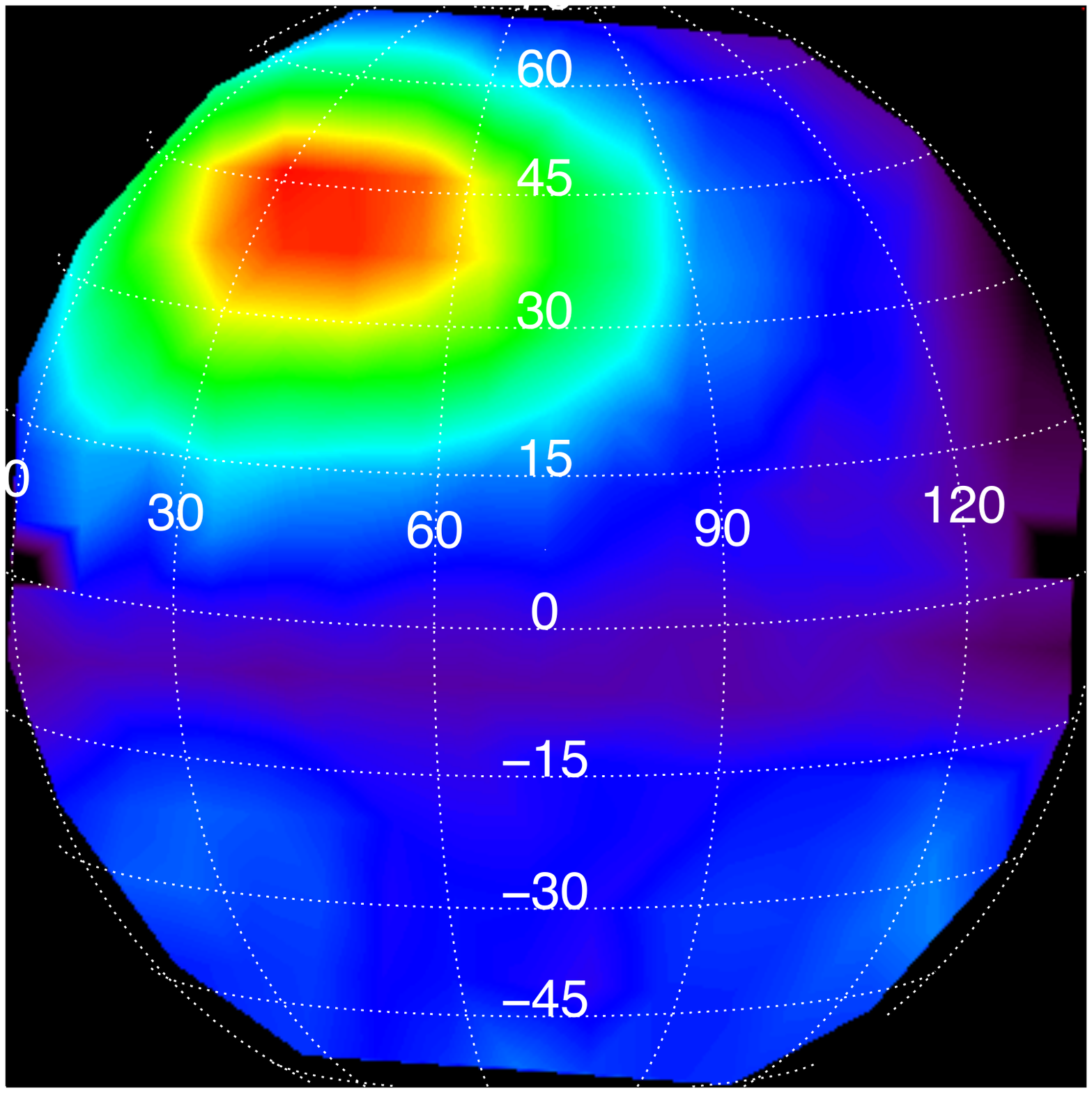}}
\subfigure{ 
\includegraphics[width=0.45\linewidth, clip=true, trim=23mm 65mm 23mm 65mm]{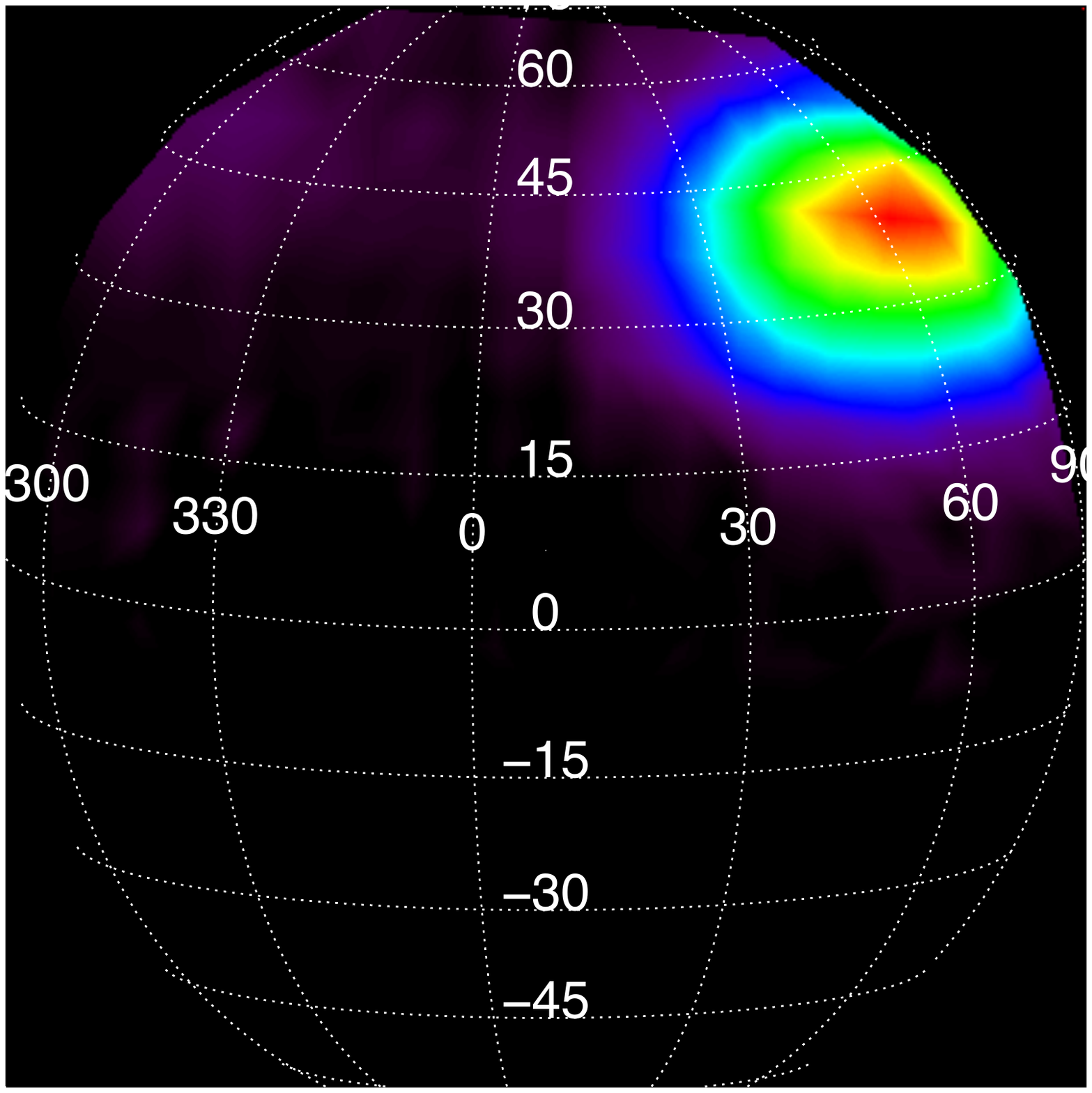}}
\caption{\label{FigScanMaps} Two examples of scan maps on July 15th orthographically projected on Saturn's disk. Left: 587-cm$^{-1}$ setting . Right: 1245-cm$^{-1}$ setting on July 15th.}
\end{figure}

The scan maps were processed with the TEXES pipeline reduction software and then post-processed through special mapping software designed to solve for the latitude and longitude on Saturn of each pixel of the scan. Two examples of these scan maps are displayed in Fig.~\ref{FigScanMaps} for the 1245-cm$^{-1}$ and 587-$cm^{-1}$ settings on July 15th. The scan maps could then be co-added into cylindrical-map projections allowing us to average much data from scan to scan to improve the signal-to-noise ratio for a given latitude and longitude on Saturn. We chose to average the data in bins of 5°$\times$5° in latitude$\times$longitude of the System III planetocentric reference system (all latitudes hereafter are given as planetocentric). Within a bin, the spectra were sorted for four different mean airmasses of 1.1, 1.35, 1.75 and 2.5. Due to the poor seeing normally found during afternoon observing on Mauna Kea, we believe we achieved at best 2\arcsec\ spatial resolution in any given map which is a sum of many scan maps. This fact makes the beacon just unresolved in latitude, but resolved in longitude (see discussion in Sec.~\ref{SecCompCIRS}).

The night of the 15th was by far the best with very little water vapor variation over time and moderate seeing. The following nights were much more difficult, with fairly large variations of water vapor and poor seeing. The variations of water vapor affected the observations of H$_2$ S(1) at 587 cm$^{-1}$ causing variations in the background transmission/emission of Earth's atmosphere. These variations occurred on timescales of seconds and thus varied throughout a single scan of Saturn, causing the continuum emission from Saturn to appear to vary over a scan map. To correct for this, we took all the data at 587~$cm^{-1}$ from a given night and summed them into a cylindrical map. Then, using the zonal average brightness of the continuum as measured from the 15th at 587 cm$^{-1}$ we added an offset to the later maps to bring their continuum levels into agreement with the data from the 15th. This is appropriate, as the reason the continuum levels disagree is due to the change in atmospheric emission from step to step in the maps caused by the telluric H$_2$O variations, an additive effect making the additive offset appropriate. Similar scaling was performed on the CH$_4$ observations, however, these observations contain regions where the continuum is very close to zero. In this case, we shifted the data in the map to make these zero regions set identically to zero.

\begin{figure}
	\includegraphics[trim=0mm 0mm 0mm 0mm, clip, width=12cm]{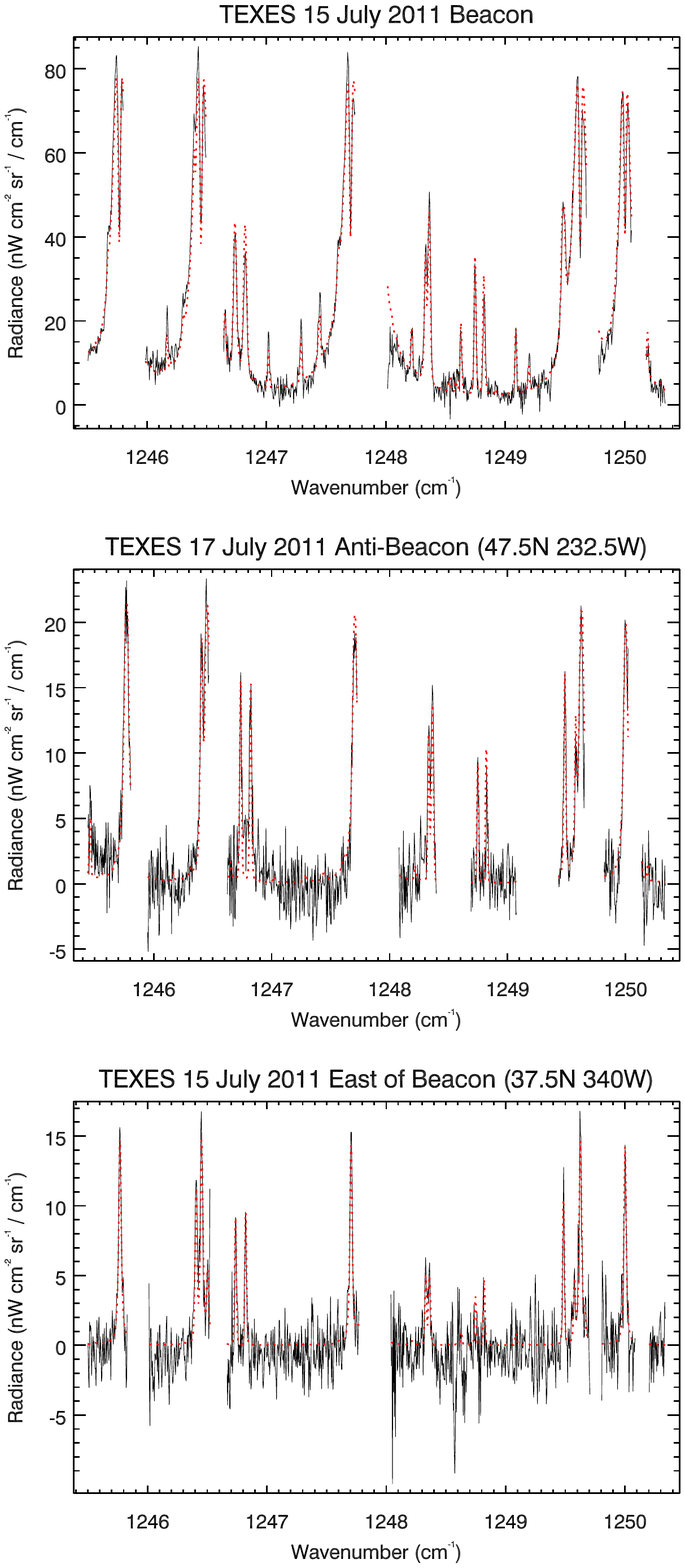}
	\caption{Three averages of TEXES spectra taken with the 1245-cm$^{-1}$ setting (black lines) compared with their best fit forward model (red lines) calculated with the temperature profiles displayed in Fig.~\ref{FigInfoCont1245}. Upper panel: Average spectrum within the beacon in the range 30°N--45°N and 30°W--55°W obtained on July 15th. Middle panel: Average spectrum in the range 40°N--55°N and 220°W--245°W obtained on July 17th. Lower panel: Average spectrum in the range 30°N--45°N and 330°W--350°W obtained on July 15th.}
	\label{Fig1245}
\end{figure}

Three examples of spectra acquired at the 1245-cm$^{-1}$ setting are shown in Fig.~\ref{Fig1245}. The upper panel displays an average of individual spectra taken on July 15th, targeting the core of the beacon between latitudes 30°N and 45°N, and longitudes 30°W and 55°W. The middle panel shows an average of spectra taken on July 17th, on the opposite side of the planet, in a warm region centered at 47.5°N and 232.5°W. The lower panel displays an average of spectra taken on July 15th, and sampling a relatively warm region situated to the east of the beacon between latitudes 30°N and 45°N, and longitudes 330°W and 350°W. Nearly all the observed emissions are due to CH$_4$, except the two weak emissions at 1248.63 and 1249.09~cm$^{-1}$, which are due to CH$_3$D. All the Saturn emissions are offset in the range $[-0.16,-0.08]$~cm$^{-1}$ with respect to their rest frequencies, in response to the $+28$~km/s velocity between Saturn and the Earth, and the $[-10,+10]$~km/s rotational velocity. As a result of this Doppler shift, absorption by the terrestrial atmosphere obscures Saturn's spectrum in the blue wing of the strongest methane lines located at 1245.75, 1246.45, 1247.70, 1249.60 and 1250.00~cm$^{-1}$. All the spectra presented hereafter have been corrected for this Doppler shift.

\begin{figure}
	\includegraphics[trim=30mm 200mm 30mm 0mm, clip, width=12cm]{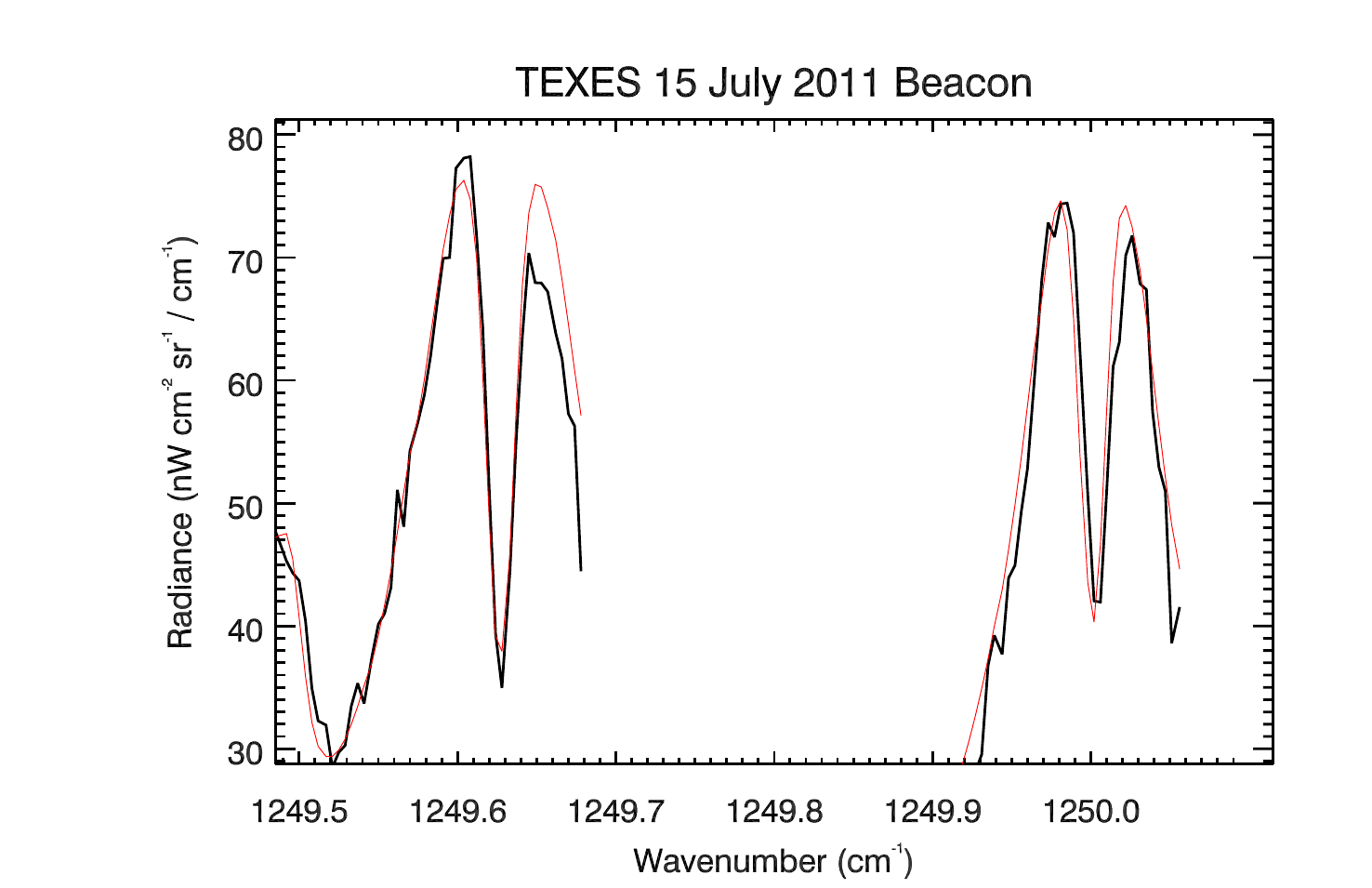}
	\caption{Zoom on two strong CH$_4$ lines from the average spectrum within the beacon in the range 30°N--45°N and 30°W--55°W obtained on July 15th and fully displayed in Fig~\ref{Fig1245}.}
	\label{Fig1245Zoom}
\end{figure}

Within the beacon, (Fig.~\ref{Fig1245}, upper panel) these strongest methane lines exhibit a broad and intense emission in their wings, and a narrow core in absorption (Fig.~\ref{Fig1245Zoom}). This line shape is archetypal of a warm atmospheric layer capped by a cold atmospheric layer, directly demonstrating that the strong atmospheric heating characteristic of the beacon was vertically limited. The weak methane lines located at 1246.17, 1247.02, 1247.29, and 1247.45 share a very high lower-energy level, $\sim$1494~cm$^{-1}$. The fact that they are observed within the beacon constitutes an evidence for very hot temperatures in this region. Outside of the beacon,  (Fig.~\ref{Fig1245}, middle and lower panels), only the methane lines with large and intermediate intensities are observed. Their line shape is radically different from that observed within the beacon, exhibiting a weak emission core and narrow wings, the line width of the spectrum taken on July 15th at 37.5°N and 340°W being narrower than that taken on July 17th at 47.5°N and 232.5°W. This line shape indicates that the warm layers in both regions were located at high altitudes, at even higher altitudes at 37.5°N and 340°W than at 47.5°N and 212.5°W.

\begin{figure}
	\includegraphics[trim=0mm 0mm 0mm 0mm, clip, width=12cm]{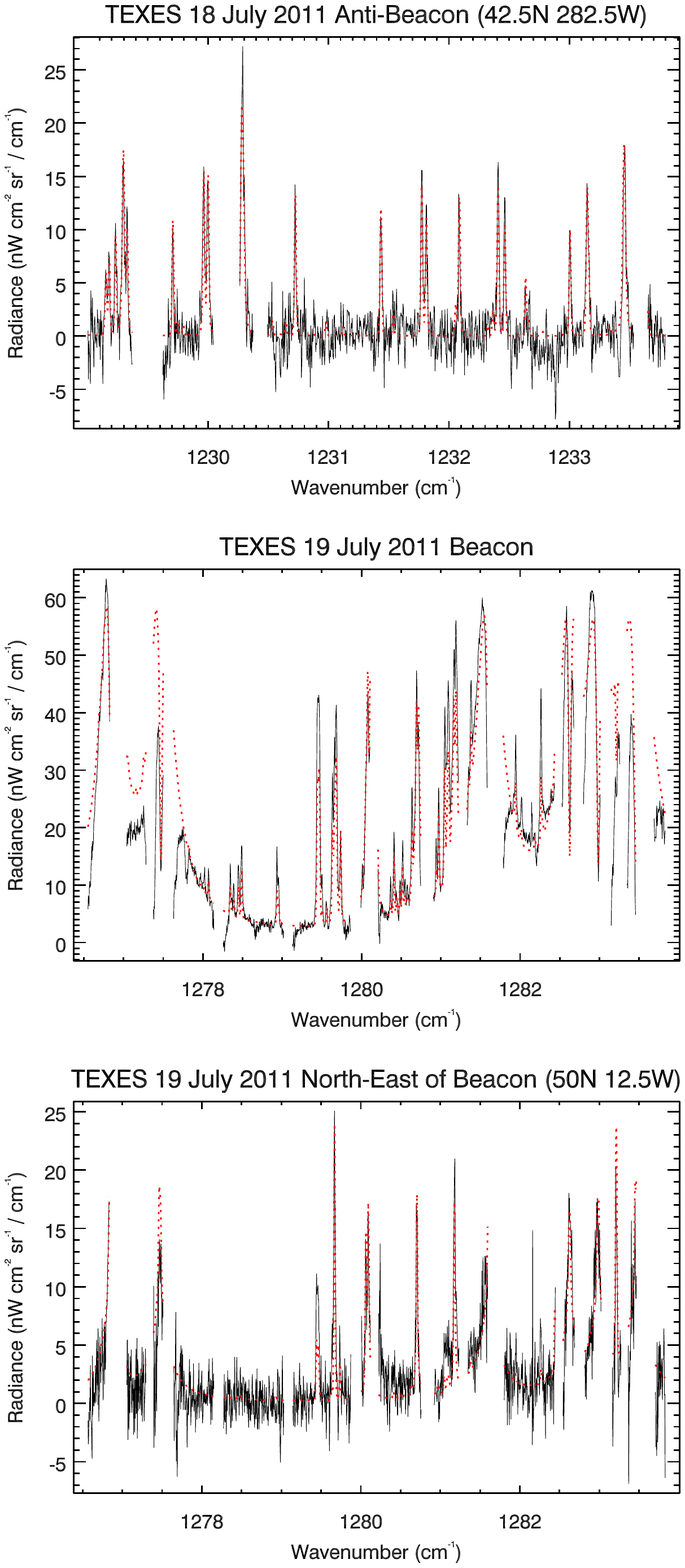}
	\caption{One average of TEXES spectra taken with the 1230-cm$^{-1}$ setting, and two averages of TEXES spectra taken with the 1280-cm$^{-1}$ setting (black lines) compared with their best fit forward model (red lines).
	Upper panel: Average 1230-cm$^{-1}$ spectrum in the range 35°N--50°N and 275°W--290°W obtained on July 18th. Middle panel: Average 1280-cm$^{-1}$ spectrum within the beacon in the range 30°N--45°N and 30°W--55°W obtained on July 19th. Lower panel: Average 1280-cm$^{-1}$ spectrum in the range 45°N--55°N and 5°W--20°W obtained on July 19th.}
	\label{Fig1280}
\end{figure}

Examples of spectra acquired at the 1230-cm$^{-1}$ setting on July 18th, and at the 1280-cm$^{-1}$ setting on July 19th are shown in Fig.~\ref{Fig1280}. The 1230-cm$^{-1}$ setting did not sample the beacon, as on July 18th it was located on the hemisphere opposite to the Earth. The upper panel of Fig.~\ref{Fig1280} hence displays an average of 1230-cm$^{-1}$ spectra opposite to the beacon, between 35°N--50°N and 275°W--290°W. In this setting, all the lines are due to methane. The middle panel displays an average of 1280-cm$^{-1}$ spectra sampling the beacon, while the lower panel displays an average of 1280-cm$^{-1}$ spectra taken to the north-east of the beacon, between latitude 45°N--55°N and longitude 5°W--20°W. Within the 1280-cm$^{-1}$ wavenumber range all the emissions are due to methane only, with the strongest lines occurring at 1276.84, 1277.47, 1280.09, 1281.61, 1282.62, and 1282.98~cm$^{-1}$. These lines exhibit the same spectral shape as the strong lines in the 1245-1250~cm$^{-1}$ wavenumber range: within the beacon, broad emission wings and a narrow absorption core, and outside of the beacon, a relatively narrow emission. This confirms the qualitative thermal vertical structure inferred from the 1245-cm$^{-1}$ setting. The beacon maximum radiance displayed in Fig.~\ref{Fig1280} also documents the lowest quality of the dataset taken on July 19th. If the spatial blurring induced by the poor observing conditions and seeing did not affect the line shape, it did affect the absolute intensities of the observed emissions. However, as for the 1245-cm$^{-1}$ setting, several weak methane lines with lower energy levels higher than 1000~cm$^{-1}$ at 1278.35--1278.50, 1278.96, 1279.45, 1279.6--1279.7, 1280.40--1280.55, 1281.0-1281.3, 1281.94, and 1282.27~cm$^{-1}$ are present in the 1280-cm$^{-1}$ beacon spectrum, demonstrating the occurrence of hot temperatures necessary to populate the energy levels of these lines.

\begin{figure}
	\includegraphics[trim=0mm 0mm 0mm 0mm, clip, width=12cm]{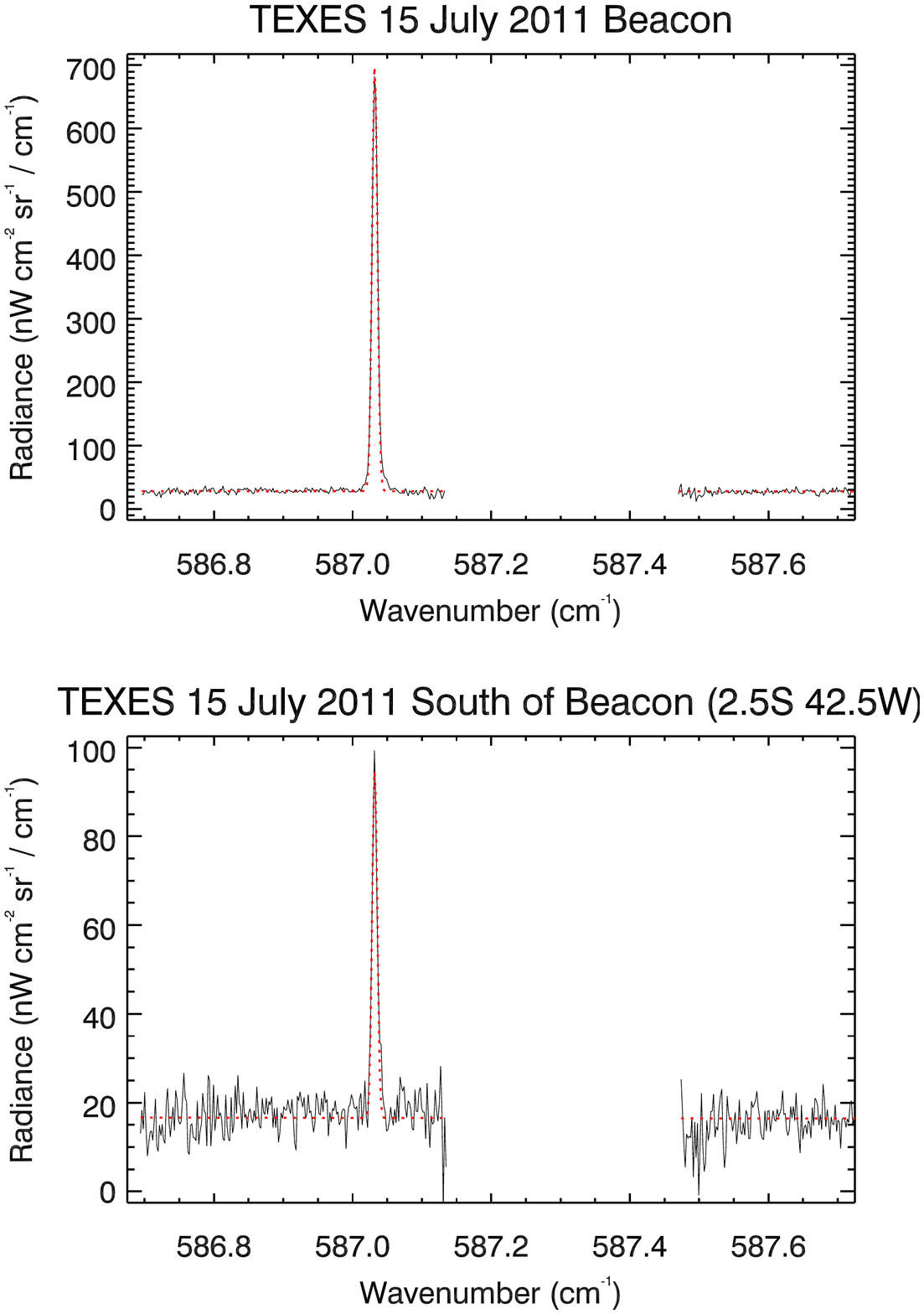}
	\caption{Two averages of TEXES spectra taken with the 587-cm$^{-1}$ setting (black lines) compared with their best fit forward model (red lines) calculated with the temperature profiles displayed in Fig.~\ref{FigInfoCont587}. Upper panel: Average spectrum within the beacon in the range 30°N--45°N and 30°W--55°W obtained on July 15th. Lower panel: Average spectrum in the range 5°S--0°N and 30°W--55°W obtained on July 15th.}
	\label{Fig587}
\end{figure}

Finally, spectra obtained at the 587-cm$^{-1}$ setting are displayed in Fig.~\ref{Fig587}, the upper panel showing an average of spectra obtained within the beacon, and the lower panel an average spectra over the equatorial region located on the beacon central meridian (between latitude 5°S and 0°S, and longitude 30°W and 55°W). In both regions, the quadrupolar S(1) H$_2$ line displays a narrow emission, extremely strong within the beacon, weak but well distinguishable from the continuum outside of the beacon. The continuum levels are nearly identical on the two regions, indicating a small meridional temperature contrast at the tropopause level. For the weak quadrupolar transitions, the Lorentz broadening at stratospheric pressures is smaller than the Doppler broadening. Hence their line profiles remain hardly unchanged throughout the stratosphere, making these lines sensitive only to one pressure level, around 1~hPa. Hence, both geographical regions exhibit the same line shape even if their inferred vertical thermal structures are drastically different. As demonstrated below in Sec.~\ref{SecRetrieval}, the quadrupolar S(1) H$_2$ actually probes the temperature at only one pressure level around 1~hPa.

\section{Data analysis}
\label{SecMethods}
\subsection{Forward radiative transfer model}

We use a standard line-by-line radiative transfer model to compute synthetic spectra for given temperature and abundance vertical profiles as described in \citet{Guerlet2009}. Our model consists of 360 layers equally spaced in $\log(\text{pressure})$ between $10^4$ and $10^{-5}$~hPa. The column density is calculated assuming hydrostatic equilibrium, taking into account the latitudinal dependence of the gravitational field due to the rapid rotation and oblate geometry of Saturn.

Our model includes the opacity due to CH$_4$, CH$_3$D, and H$_2$ quadrupolar lines and collision-induced continuum. For CH$_4$ and CH$_3$D, we compute the opacity using the  spectroscopic line parameters from the GEISA 2011 database \citep{JacquinetHusson2008}. For the H$_2$ quadrupolar lines we use the \citet{Campargue2012,Hu2012} intensities and positions, and the broadening parameters from \citet{Reuter1994}. Finally, we take into account the H$_2$ collision-induced continuum due H$_2$, He and CH$_4$ collision partners using the algorithm and numerical values of \citet{Borysow1986} and \citet{Borysow1985,Borysow1988}. 

For the molecular abundances, we use the deep volume mixing ratio of \citet{Flasar2005} for CH$_4$, $4.5\times10^{-3}$, inferred from CIRS spectra of Saturn, the deep volume mixing ratio of \citet{Lellouch2001} for CH$_3$D, $3\times10^{-7}$, inferred from Infrared Space Observatory (ISO) data. The vertical variation of the methane abundance due to photolysis and the prevalence of molecular diffusion at high altitudes is taken into account following the work of \citet{Moses2000}. The relative abundance of He to H$_2$ is still  poorly constrained in Saturn's atmosphere. For consistency with the CIRS limb spectra analysis presented by \citet{Guerlet2009,Guerlet2013,Sylvestre2015}, we use a constant volume mixing ratio of 0.86 for molecular hydrogen and 0.1355 for helium, values similar to the accurate measurements obtained by \citet{vonZahn1998} for Jupiter from the Galileo \textit{in situ} soundings, and lying within the proposed range for Saturn by \citet{Conrath2000} from a reanalysis of Voyager/IRIS data. The sensitivity of our retrieved temperatures to the assumed helium abundance will be presented in Sec.~\ref{SecErrors}.

\subsection{Temperature inversion}
\label{SecRetrieval}

The retrieval of a temperature vertical profile from spectroscopic observations is a classical ill-posed problem. To solve this problem, we used a constrained and regularized retrieval algorithm following the method detailed in \citet{Conrath1998} and \citet{Rodgers2000}. With such a method, the retrieved temperature profiles were constrained to stay close to an \textit{a priori} profile $\mathbf{T}_0$ at pressures where the measurements have no information, and the departure from the \textit{a priori} profile elsewhere was regularized, or smoothed, to inhibit strong vertical oscillations.

The algorithm starts by linearizing the dependence of the observed radiance with the temperature
\begin{equation}
\Delta I_i =\sum_{j=1}^n \frac{\partial I_i}{\partial T_j}\Delta T_j
\label{EqLinear}
\end{equation}
where $\Delta I_i$ is the difference between the observed and synthetic radiance at a wavenumber $\nu_i$, and $\Delta T_j$  is the difference, at the pressure level $p_j$, between the actual value of the temperature and the temperature used in the forward radiative model. Introducing the Jacobian matrix $K$
\begin{equation}
K_{ij}= \frac{\partial I_i}{\partial T_j}
\end{equation}
and the vectors $\mathbf{\Delta I}$ and $\mathbf{\Delta T}$, Equation~\ref{EqLinear} can be rewritten in a vectorial form
\begin{equation}
\mathbf{\Delta I} = K \mathbf{\Delta T}
\end{equation}

The formal solution to the ill-posed problem is
\begin{equation}
\mathbf{\Delta T} = U\mathbf{\Delta I} \quad \text{with}\quad U=\alpha SK^T(\alpha KSK^T + E)^{-1}
\end{equation}
where we introduce the scalar $\alpha$, the weight with which the constraint relative to the least squares fitting of the observation is imposed to the solution, the correlation matrix $S$ regularizing the solution, and the error covariance matrix of the measurement $E$ (see Sec.~\ref{SecErrors} for its determination). In our analysis, we set the correlation matrix $S$ to be a simple Gaussian function with a correlation length $L$ equal to an atmospheric scale height.  We also found that the best values of the parameter $\alpha$ was obtained when it equaled the traces of the matrices $E$  and $\alpha KSK^T$.

As the Planck function and the atmospheric opacity do not vary linearly with the temperature, the algorithm has to proceed by successive iterations: starting from the \textit{a priori} temperature vertical profile $\mathbf{T}_0$, the inversion process is run several times, with the temperature profile $\mathbf{T}_n$ obtained at the $n$th iteration being used to calculate the Jacobian matrix and the synthetic spectrum of the $(n+1)$th iteration. In our analysis, the iteration was stopped when the quality of the fit, estimated by  the least square $\chi^2=\mathbf{\Delta I} E^{-1} \mathbf{\Delta I}^T$, changed from less than $1\%$ from an iteration to another.

The averaging kernel $A$ is an important quantity to estimate the information content of the measurement
\begin{equation}
A = UK
\end{equation}
The rows $\mathbf{a}_j^T$ of the matrix $A$ are functions of pressure, whose area represents the relative weight between the fraction of the retrieved temperature $T_j$ that comes from the measurement to the fraction that comes from the \textit{a priori} profile, whose half-width represents the vertical resolution of the retrieved temperature profile, and whose peak determines the pressure of maximum sensitivity to the temperature. If the peak of a function $\mathbf{a}_j^T$ corresponds to a pressure level $p_{j'}$ different from the pressure level $p_j$, it means that the temperature $T_j$ at the pressure level $p_j$ is not determined independently from the measurement, but rather from the \text{a priori} profile and the correlation with the retrieved temperature at the pressure level $p_{j'}$. It follows immediately that the averaging kernel $A$ determines the vertical range probed by the observations. The number $d$ of independent temperatures retrieved from the observations is given by the expression
\begin{equation}
d=\Tr(A)
\end{equation}

\begin{figure}
	\includegraphics[trim=0mm 25mm 0mm 15mm, clip, width=15cm]{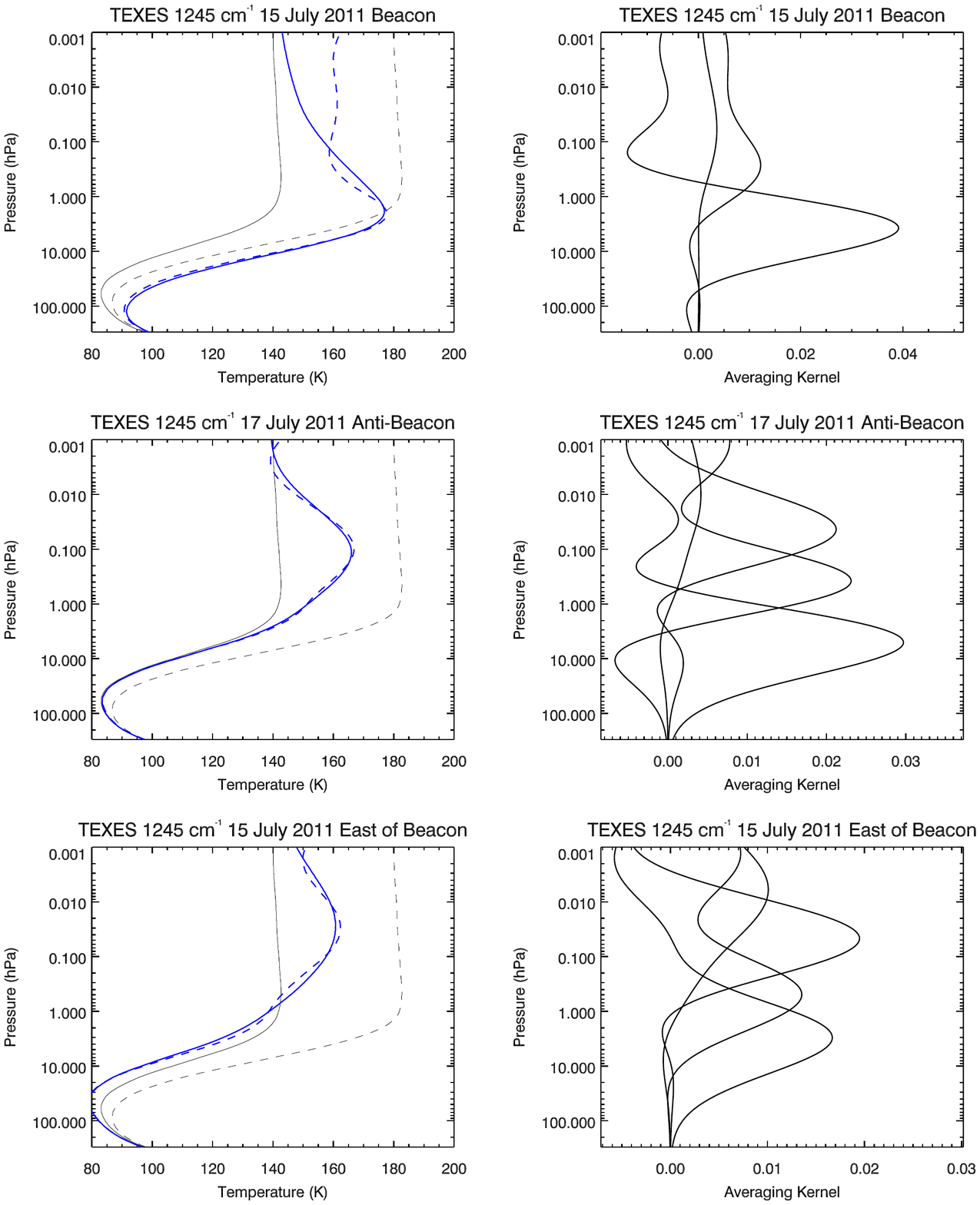}
	\caption{Left column: The inverted temperature profiles (blue lines) for the three average spectra presented in Fig.~\ref{Fig1245}, and the \textit{a priori} profiles (black lines) for the \citet{Lindal1985} profile (solid line) and the 40K warmer profile (dashed line). Right column: The averaging kernels yielded by the inversion algorithm for the same three average spectra, at four pressure levels (3 hPa, 0.4 hPa, 0.06 hPa, and 0.01 hPa), except within the beacon, where only the first three pressure levels are shown. }
	\label{FigInfoCont1245}
\end{figure}

The \textit{a priori} profile plays an important role in the inversion process. Outside the sensitivity region of the measurement, the inverted profile relaxes toward the \textit{a priori} profile, where its actual value may still affect slightly the retrieved values within the range of maximum sensitivity. An \textit{a priori} profile very different from the final profile may also produce spurious oscillations. In our analysis, we used two different \textit{a priori} temperature profiles. The first one, the \textit{cold} one, was based on the profile inferred by \citet{Lindal1985} from Voyager radio occultation that we smoothed to remove structures at vertical scales smaller than one scale height and relaxed toward an isothermal atmosphere above the 0.1-hPa pressure level. But we also investigated a warmer \textit{a priori} profile, smoothly departing from the \citet{Lindal1985} profile at the tropopause level to differ by 40K above the 1-hPa pressure level (Fig.~\ref{FigInfoCont1245}).

Figure~\ref{FigInfoCont1245} displays the results of the inversion algorithm for the three average spectra obtained with the 1245-cm$^{-1}$ setting presented in Fig.~\ref{Fig1245}. The left hand side column presents the inverted profiles (blue lines) and the \textit{a priori} profiles (dark lines) for the \citet{Lindal1985} profile (solid line) and the 40K warmer profile (dashed line). The right hand side column presents the averaging kernels at four pressure levels (3 hPa, 0.4 hPa, 0.06 hPa , and 0.01 hPa), except within the beacon, where only the first three pressure levels are shown.  Indeed, inspecting the upper row of Fig.~\ref{FigInfoCont1245} for this warm region, it is clear that our maximum sensitivity is limited to the 20--0.1 hPa pressure range. Above the 0.1-hPa pressure level, the averaging kernels vanish to non-significant values. It results from this situation that the two profiles inverted from the two different \textit{a priori} temperature profiles disagree above the 0.1-hPa pressure level. The numerical calculation states that 2.5 independent temperatures can be measured. They correspond to the strong hot peak at 2~hPa, and to the distinctly cooler region above this level, already inferred from the shape of the methane lines. The remaining 0.5 degree of freedom given by the inversion accounts for the fact that, above the 0.1-hPa  pressure level, the temperature profile inverted from the warm \textit{a priori} does not relax towards 180K, demonstrating that the temperature must be lower than 160K between 0.1~hPa and $10^{-3}$~hPa.

Outside of the beacon, our sensitivity extends to lower pressures than within the beacon. The averaging kernels have significant values between 5~hPa and $5\times10^{-3}$~hPa, the temperature profiles inverted from both \textit{a priori} profiles agree up to the $2\times10^{-3}$-hPa pressure level, and the number of degrees of freedom given by the algorithm is close to 3. As already pointed in Sec.~\ref{SecObs}, the spectra located to the west of the beacon yield a maximum temperature at the 0.1-hPa pressure level, while the spectra located to the east of the beacon yield a maximum temperature centered a scale height higher, at the 0.03-hPa pressure level.

\begin{figure}
	\includegraphics[trim=0mm 0mm 0mm 50mm,  width=15cm,angle=90]{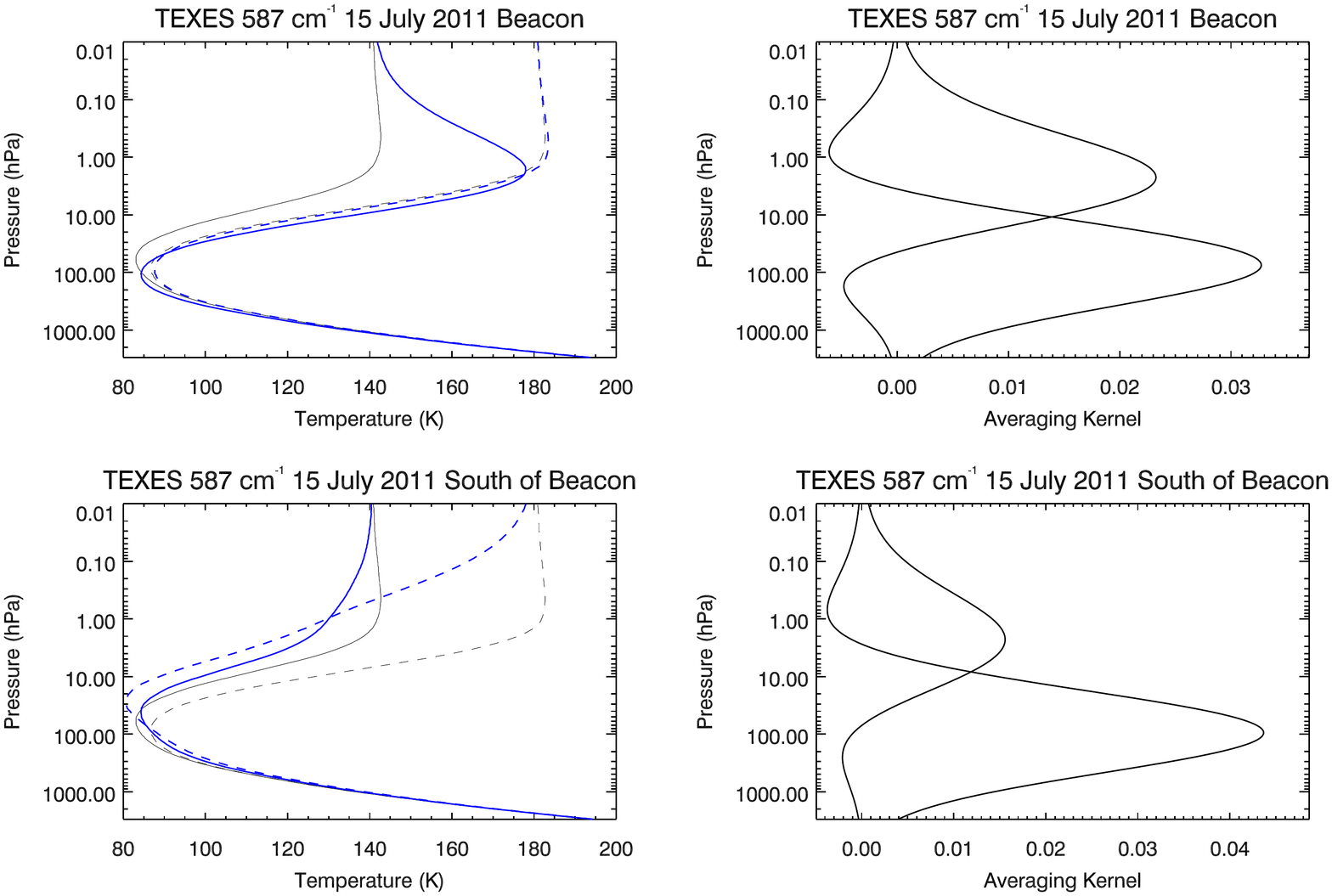}
	\caption{Left column: The inverted temperature profiles (blue lines) for the two average spectra presented in Fig.~\ref{Fig587}, and the \textit{a priori} profiles (dark lines) for the \citet{Lindal1985} profile (solid line) and the 40K warmer profile (dashed line). Right column: The averaging kernels yielded by the inversion algorithm for the same two average spectra, at two pressure levels (100 hPa, and 2 hPa).}
	\label{FigInfoCont587}
\end{figure}

Figure~\ref{FigInfoCont587} displays the results of the inversion algorithm for the two average spectra obtained with the 587-cm$^{-1}$ setting presented in Fig.~\ref{Fig587}.  Two independent pressure levels are probed by this setting, the 1--2 hPa pressure by the quadrupolar line, and the 80--100 hPa pressure by the collision-induced absorption. Between the tropopause level and the 1--2 hPa level, the temperature is left unconstrained. Above the 1-hPa pressure level, the inverted profile quickly relaxes towards the \textit{a priori}, leading to sharp differences between the profiles inverted from the cold and the warm \textit{a priori} temperature profiles.

\subsection{Error Analysis}
\label{SecErrors}

In our error analysis we distinguished the random uncertainties that affected the precision of our measurements, and the systematic uncertainty that affected the accuracy of our results. Here, we present first the estimation of the precision of our measurements, then the accuracy of our measurements.

The first source of uncertainty that affects our retrieval is the Noise Equivalent Spectral Radiance (NESR). For the CH$_4$ settings, it can be readily estimated for spectra obtained outside of the beacon between methane lines where the expected and observed radiances are null. The same estimation cannot be performed for spectra obtained within the beacon, since the radiance between lines never reaches the zero level, due to the extreme extent of the emission coming from the strong CH4 line wings. For these spectra, we estimated the NESR from spectra obtained south of the beacon for the same slit position on the planet. In the case of the H$_2$ spectra, the NESR can be estimated as the mean standard deviation of the radiance in the smooth and broad continuum away from the S(1) quadrupolar line. The second source of uncertainty is the error in the telluric transmission correction. To account for this error, we weighted the NESR by the factor $T(\nu)/\sqrt{1.1-T(\nu)}$, where $T(\nu)$ is the telluric transmission at the wavenumber $\nu$ \citep{Greathouse2005}.
A third source of uncertainty is the smoothing error, which is an error linked to the limited vertical resolution of the observations. Indeed, the averaging kernels displayed in Fig.~\ref{FigInfoCont1245} and Fig.~\ref{FigInfoCont587} have a typical FWHM of one scale height. The combination of these three uncertainties cumulate to an error on the retrieved temperature profile given by the expression:
\begin{equation}
\sigma=  S - SK^T (KSK^T - E )^{-1} KS
\end{equation}
The retrievals using the CH$_4$ lines have a precision, or random error, of 0.5--1K at 1~hPa depending on the quality of the spectra and 1--2K in the pressure range 0.1--0.01 hPa, while the retrievals using the H$_2$ lines have a precision 0.3--0.5K at 100~hPa, and 0.5--1K at 1~hPa.

Moreover, two systematic effects limit the accuracy of our temperature measurements. The first systematic effect comes from the uncertainties on the molecular hydrogen and methane abundances. Hydrogen volume mixing ratio (vmr) was proposed to lie in the range 0.84--0.89 by \citet{Conrath2000}, but a more recent analysis combining CIRS observations and radio occultations \citep{Flasar2008} suggested that the hydrogen vmr could be as high as 0.92. On the methane side, the studies of \citet{Flasar2005} and \citet{Fletcher2009} showed that its abundance is known to within $\pm5\%$. To calculate the accuracy of our temperature measurements implied by these composition uncertainties, we ran our model using the two 0.84 and 0.92 extreme hydrogen abundances, and using methane profiles that differed from our nominal profile by $\pm5\%$. The differences between the corresponding retrieved temperature profiles and the nominal inverted profile set the accuracy of our measurements. At 100~hPa, the retrieved temperature would have been 1.5K colder using the 0.92 H$_2$ vmr rather than our nominal value of 0.86, and 0.5K warmer using the 0.84 H$_2$ vmr respectively. A 1~hPa, the impact of the He uncertainty is lower, lying in the range $[-0.5,+0.2]$K within the beacon, and $[-0.3,+0.1]$K outside of the beacon. For CH$_4$, the uncertainties on its abundance limit the  accuracy of our measurements to $\pm0.4$K in the pressure range 0.01--1~hPa outside of the beacon, while inside the beacon the accuracy decreases to $\pm1$K at the 1~hPa level, and $\pm0.5$K at lower pressures.
 
The second systematic effect comes from the uncertainty on TEXES absolute calibration, estimated to lie within $\pm20\%$. As for estimating the accuracy limit due to the uncertainties on Saturn's composition, we ran our inversion algorithm on CH$_4$ and H$_2$ spectra multiplied by a factor of 0.8 and 1.2. For the temperature retrieved from CH$_4$ spectra, the accuracy limit rises from $\pm2$K outside of the beacon up to $\pm4$K within the beacon. For the temperature retrieved from H$_2$ spectra, the accuracy changes from $\pm5$K to $\pm7$K at 1~hPa respectively outside and within the beacon, while at 100~hPa, the accuracy is of the order of $\pm2$K.

In the following presentation of the results, we will quote only the precision of the measurements, but the reader must keep in mind that the results maybe systematically offset by a few Kelvin.

\section{Results and comparison with previous measurements}
\label{SecResults}

\subsection{Horizontal thermal structure}

\begin{figure}[ht!]
\subfigure{ 
\includegraphics[width=0.45\linewidth, clip=true, trim=00mm 0mm 0mm 0mm]{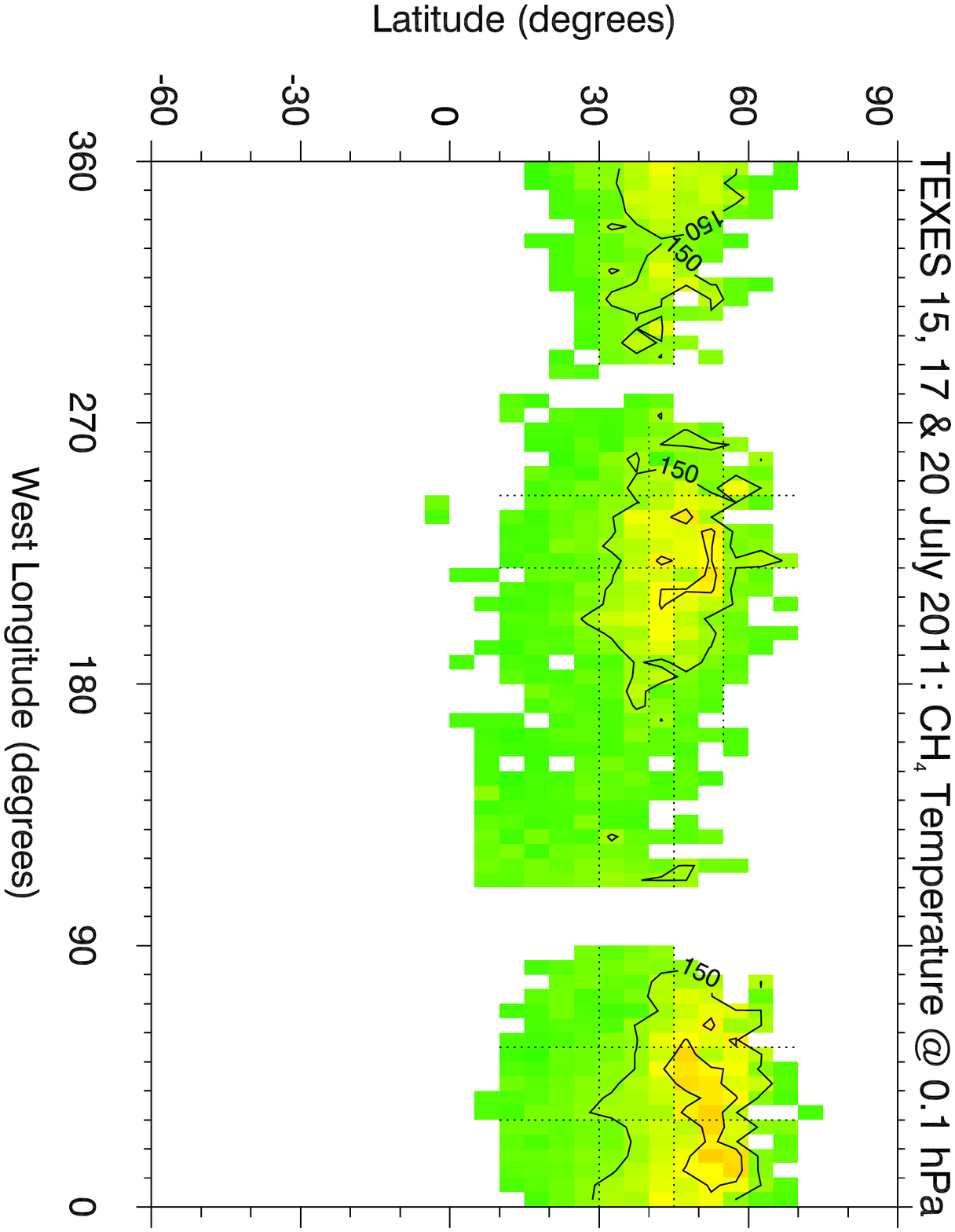}
\subfigure{ 
\includegraphics[width=0.45\linewidth, clip=true, trim=00mm 0mm 0mm 0mm]{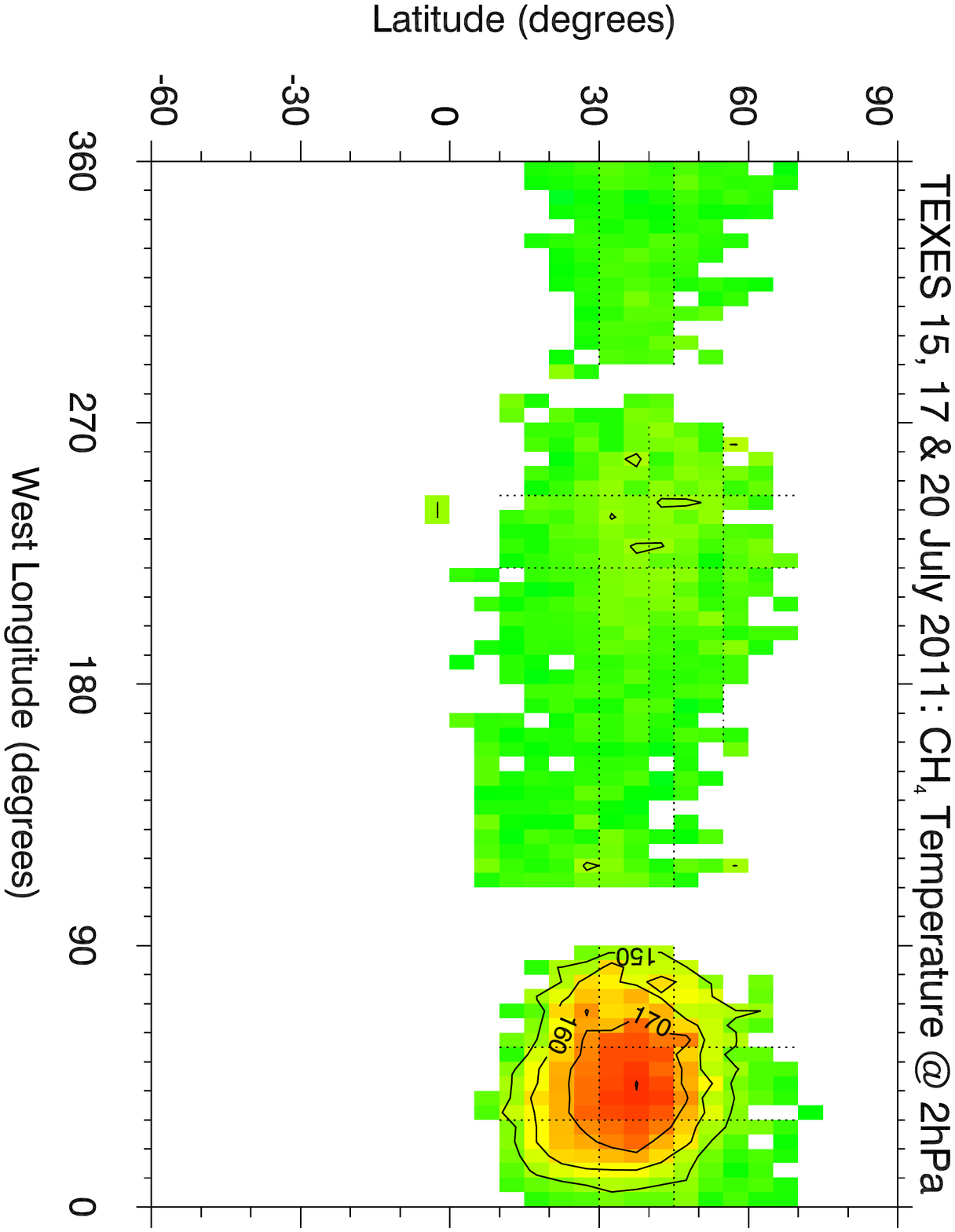}}
\label{FigTempMap1245at2hPa}}
\subfigure{ 
\includegraphics[width=0.45\linewidth, clip=true, trim=00mm 0mm 0mm 0mm]{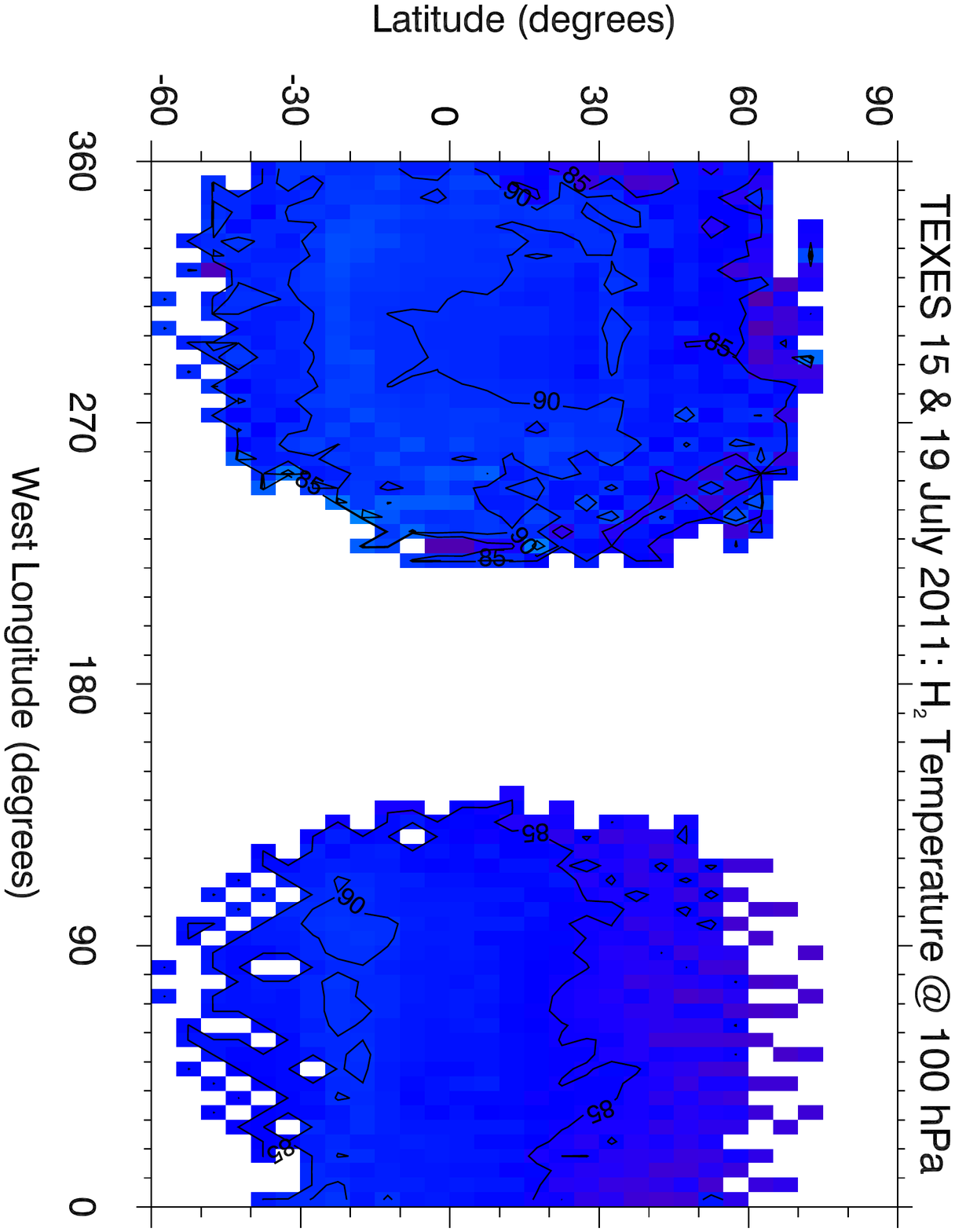}}
\subfigure{
\includegraphics[width=0.45\linewidth, clip=true, trim=00mm 0mm 0mm 0mm]{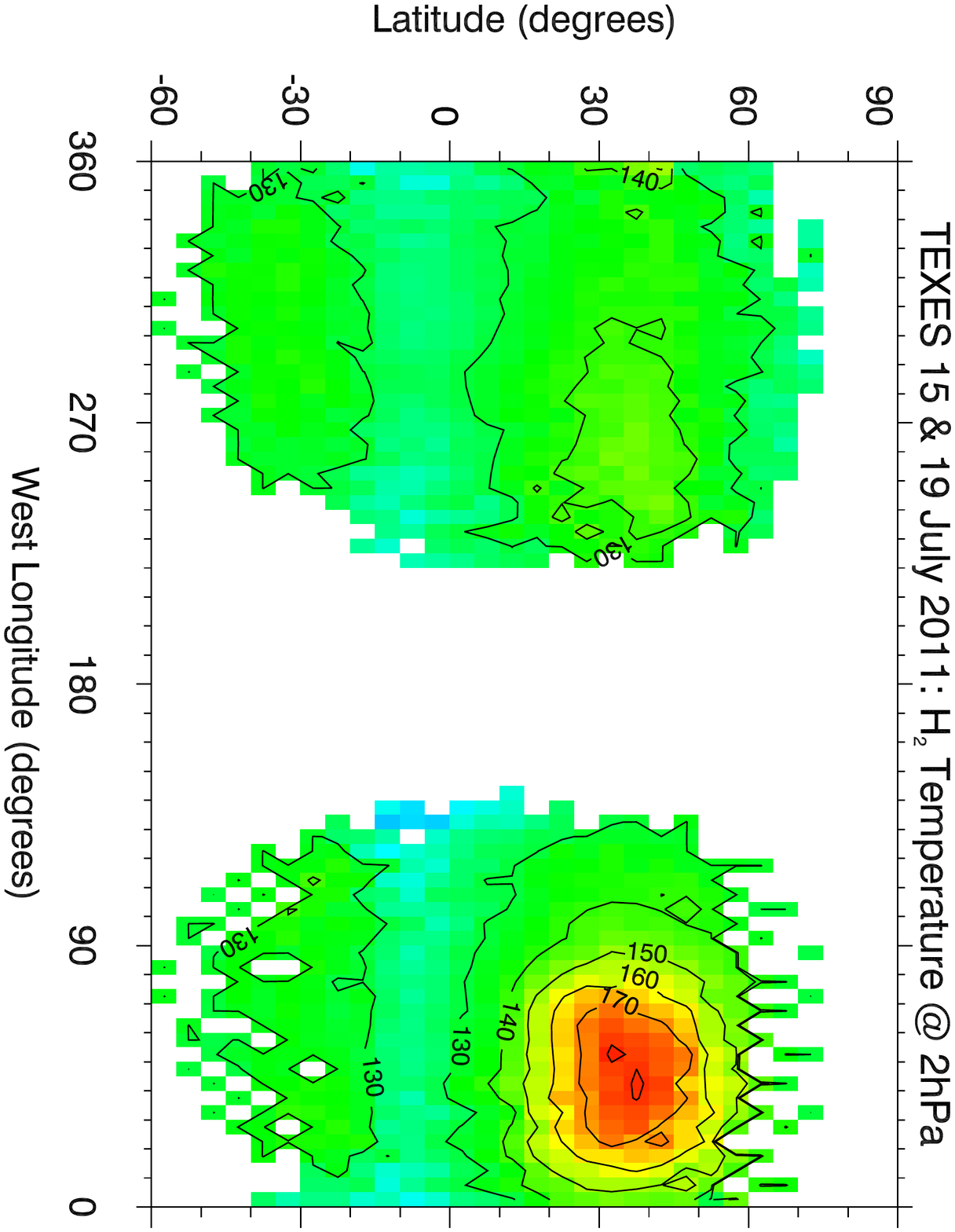}}
\caption{\label{FigTempMap} Maps of temperatures at 2 hPa (upper left) and 0.1 hPa (lower left) inferred from the 1245-cm$^{-1}$ setting on  July 15th, 17th, and 20th, and at 2 hPa (upper right) and 100 hPa (lower right) inferred from the 587-cm$^{-1}$ setting on July 15th and 19th. The dotted lines show the limits of the zonal and meridional averages made to obtain the latitude-pressure and the longitude-pressure cross sections of the temperature presented in Fig.~\ref{FigCrossSections}.}
\end{figure}

\begin{figure}[ht!]
\subfigure{ 
\includegraphics[width=0.6\linewidth,angle=90]{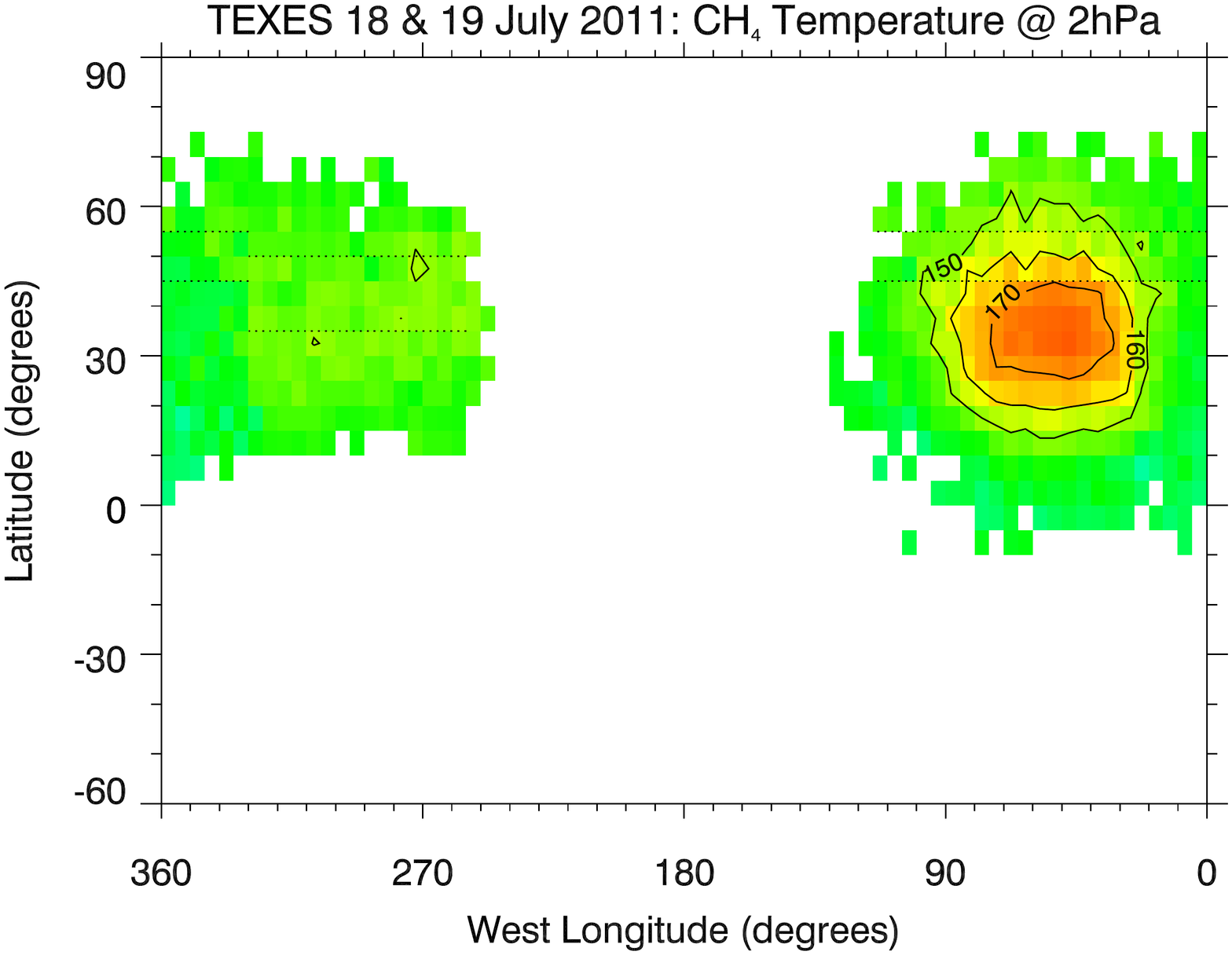}}
\subfigure{ 
\includegraphics[width=0.6\linewidth,angle=90]{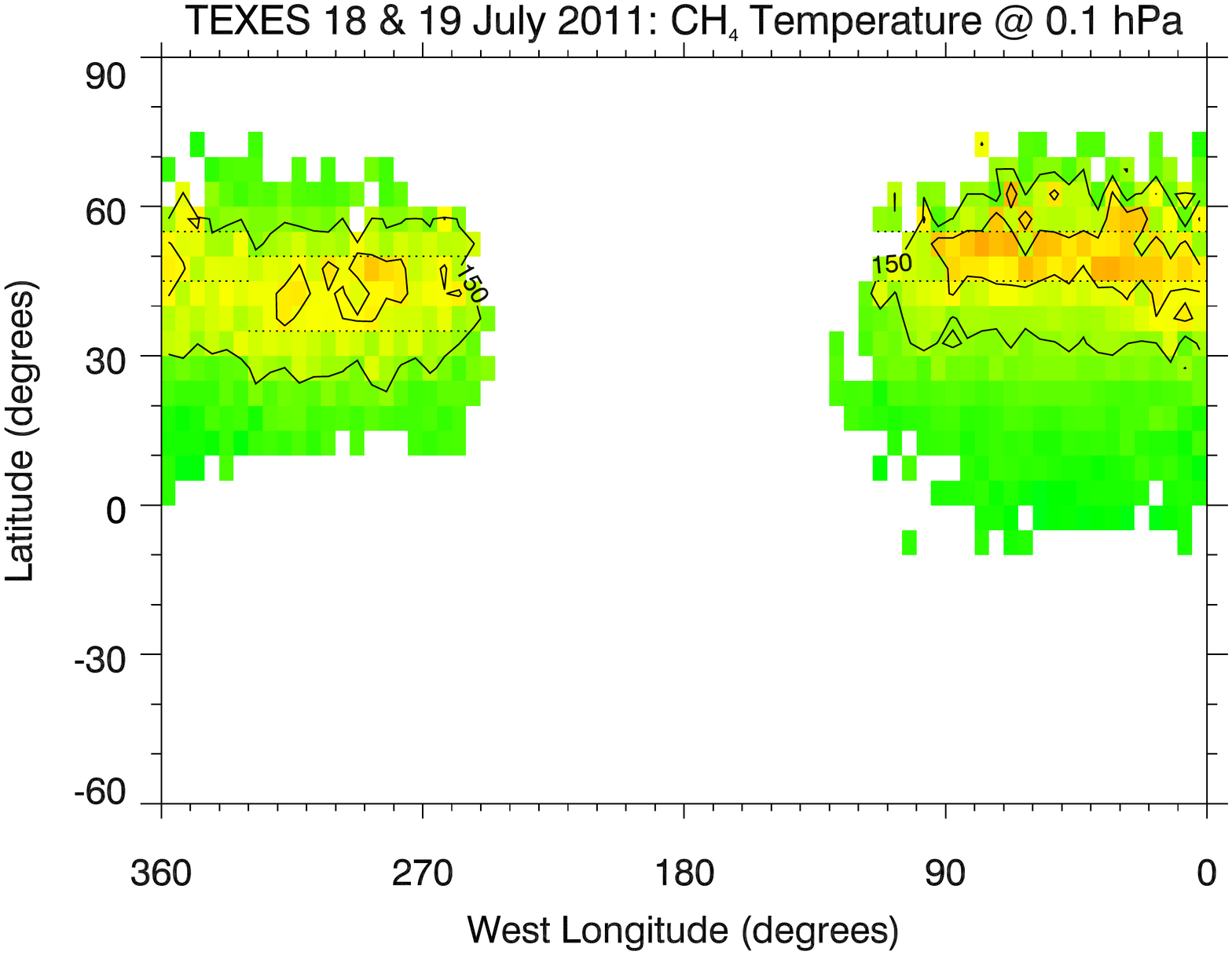}
\label{FigTempMap1280at2hPa}}
\caption{\label{FigTempMap19} Maps of temperatures at 2 hPa (upper panel) and 0.1 hPa (lower panel) inferred from both the 1230-cm$^{-1}$ setting on  July 18th and the 1280-cm$^{-1}$ setting on  July 19th. The dotted lines show the limits of the zonal and meridional averages made to obtain the latitude-pressure and the longitude-pressure cross sections of the temperature presented in Fig.~\ref{FigCrossSections}.}
\end{figure}

We first retrieved Saturn's  thermal structure by coadding all TEXES spectra taken at similar airmasses in boxes of 5°$\times$5° in latitude$\times$longitude as explained in Sec.~\ref{SecObs}, and by inverting these spectra with the algorithm presented in Sec.~\ref{SecRetrieval}. Figure~\ref{FigTempMap} presents cylindrical projection maps of the retrieved temperature from the 1245-cm$^{-1}$ spectra taken on July 15th, 17th, and 20th (left column) at the 2-hPa and 0.1-hPa pressure levels, and from the 587-cm$^{-1}$ setting on July 15th and 19th (right column) at the 100-hPa and 2-hPa pressure levels. Figure~\ref{FigTempMap19} presents temperature maps retrieved from both the 1230-cm$^{-1}$ setting on July 18th and the 1280-cm$^{-1}$ setting on July 19th, at the 2-hPa and 0.1-hPa pressure levels.

At the 2-hPa pressure level (upper rows of Fig.~\ref{FigTempMap} and \ref{FigTempMap19}), the warm beacon oval is evident in the two temperature maps retrieved using the CH$_4$ lines, as well as from the temperature map retrieved from the H$_2$ line. It appears as a warm anomaly of $\sim$30K--35K, sitting on a quiescent background temperature of 145$\pm1$K, with a FWHM of 60°--70° in longitude and 30° in latitude. The  1245-cm$^{-1}$ and 587-cm$^{-1}$ settings indicate that the beacon was centered at 37.5°N and 42.5°W on July 15th, while the 1280-cm$^{-1}$ setting shows it was centered at 37.5°N and 52.5°W on July 19th. The maximum temperatures inferred from the CH$_4$ 1245-cm$^{-1}$ and 1280-cm$^{-1}$ settings are 180$\pm0.5$K and 176$\pm0.5$K respectively, while it peaks at 181.5$\pm0.5$K for the temperature retrieved from the H$_2$ line. The good agreement between the temperatures retrieved from the CH$_4$ lines and the H$_2$ S(1) line on the same date (July 15th), much better than the relative accuracy of the two measurements, indicates that an accurate relative calibration between the two settings was achieved by using the room temperature black body calibration source.

Our measured beacon positions and widths are in agreement with that retrieved by \citet{Fletcher2012} from Cassini/CIRS spectra and VLT/VISIR images. These authors found a longitudinal FWHM of 70--80°, a latitudinal FWHM of 30--35°, and a center latitude of 30°N--35°N. Our dates of observation, from July 15th to July 20th, correspond to the transition between Phase II and Phase III of the beacon evolution, as identified by \citet{Fletcher2012}, when the beacon longitude drift rate with respect to System III accelerated from (1.6°$\pm$0.2°)/day to (2.7°$\pm$0.04°)/day. The transition between the two drift rates was relatively sharp; it occurred between July 8th, when \citet{Fletcher2012} measured a center position of 30°W, and  July 26th when the beacon was centered at 70°W. The difference between our center longitude of 42.5°W measured on July 15th and the beacon center measured by CIRS on July 8th, is indeed compatible with a drift rate  of 1.6°/day. In contrast, our center position of 52.5°W measured on July 19th, is compatible with a drift rate 2.5°/day between July 15th and July 19th, as well as between the TEXES observations performed on July 19th and the CIRS sequence taken on July 26th. Hence, our observations suggest that the transition between Phase II and Phase III could have been even sharper than revealed by the CIRS dataset, occurring in the July 15th--July 19th interval. However, a definitive conclusion cannot be reached on this issue because of our 5°-uncertainty on the exact beacon center longitude.

The background temperature of 145$\pm1$K we measure outside of the beacon at the 2-hPa pressure level from the TEXES dataset is identical, within the error bars, to the temperature measured from CIRS limb measurements by \citet{Sylvestre2015} on September 23rd, 2010, just before the onset of the GWS, or from CIRS nadir spectra on July 8th \&\ 26th, 2011 by \citet{Fletcher2012}. However, our inferred maximum temperature within the beacon is significantly lower than that measured by \citet{Fletcher2012} from the CIRS dataset. From observations taken on July 8th, they inferred temperatures in excess of 200K, while on July 26th, they found a maximum temperature still warmer than 190K. This disagreement between the CIRS and TEXES retrieved temperatures is specifically addressed in Sec.~\ref{SecHotLines} and \ref{SecCompCIRS}. The difference in maximum temperature measured from the TEXES dataset on July 15th and July 19th, respectively 180$\pm0.5$K and 176$\pm0.5$K, already suggests that atmospheric blurring of TEXES spectra is the major reason for this mismatch. 

At 0.1~hPa, the temperature maps inferred from the 1245-cm$^{-1}$ setting on July 15th, 17th  \&\ 20th (Fig.~\ref{FigTempMap}, lower left panel), and the 1230 \&\ 1280-cm$^{-1}$ setting on July 18th \&\ 19th (Fig.~\ref{FigTempMap19}, lower panel) are radically different from that retrieved at the 2-hPa pressure level on the same dates. Two main temperature anomalies can be observed. The first one is located north of the beacon, between latitudes 45°N--60°N with a maximum temperature centered at 52.5°N, and extends to a wider longitudinal area than the beacon on its eastern and western sides, between 340°W and 90°W. In this region, the maximum temperature is measured at 165$\pm1$K on July 15th, and at 168.5$\pm1$K on July 19th. The second anomaly is located to the east of the beacon, stretching from 330°W to 190°W, nearly opposite to the beacon, and latitudes 35°N and 50°N. The peak temperature of 163$\pm1$K is located at 52.5°N and 217.5°W. This eastward warm anomaly contrasts with the situation to the west of the beacon, where the temperature is relatively uniform in the range 145-150K, albeit sparsely sampled. We note that the temperature field in the 270°-330°W longitude range at this 0.1-hPa pressure level measured using the 1245-cm$^{-1}$ setting on July 15th does not exactly agree with that measured from the 1230-cm$^{-1}$ setting on July 18th, the former giving lower temperatures than the latter. As this longitudinal range was sitting close to the limb of planet on July 15th, we favor the temperatures retrieved on July 18th, when the sub-Earth longitude ranged between 288°W and 341°W.

It is difficult to compare our temperature maps obtained at the 0.1-hPa pressure level with the measurements presented by \citet{Fletcher2012} as CIRS nadir spectra are only marginally sensitive to the temperature at pressures lower than 0.5~hPa. Nevertheless, we note that the longitude-pressure cross section obtained from CIRS spectra on July 8th shows warmer 0.5-hPa temperatures on the eastern edge of the beacon than on its western edge, with a decrease of the temperature from 360°W to a minimum at 160°W. The temperature map at 0.5~hPa obtained from CIRS on August 21st also show this warm tail, slightly to the north of the beacon, in agreement with our TEXES observations. Still, it is more relevant to compare our measured temperature anomalies at 0.1~hPa with CIRS limb measurements whose spherical geometry allowed \citet{Guerlet2009,Guerlet2013} and \citet{Sylvestre2015} to measure temperature vertical profiles up to the 0.01-hPa pressure level. Before the storm outbreak, \citet{Sylvestre2015} retrieved temperatures of 143$\pm2$K between 30°N and 50°N, slowly dropping to 140$\pm2$K at 55°N and 135$\pm2$K at 77.5°N. These temperatures measured in September 2010 had only marginally increased since the previous CIRS limb observation campaign that took place in 2004--2005, from which \citet{Guerlet2009} had measured temperatures in the range 138--142K. After the storm, \citet{Guerlet2013} measured temperatures in the range 158K--165K at 0.1~hPa on the east side of the beacon from three observations acquired between July 10th, 2011 and August 23rd, 2011. Hence, both TEXES and CIRS limb spectra point to a temperature rise of 20K at the 0.1-hPa pressure level in less than a year between September, 2010 and July, 2011. Such an increase cannot be attributed to the radiative seasonal evolution of the stratospheric thermal structure \citep{Guerlet2014,Sylvestre2015}, but must be linked to the beacon and the stratospheric aftermath of the 2010 GWS. This difference between CIRS and TEXES temperature will be specifically addressed in Sec.~\ref{SecCompCIRS}.

At the 100-hPa pressure level, the TEXES temperature map obtained from the 587-cm$^{-1}$ setting on July 15th and July 19th is shown on the lower right panel of Fig.~\ref{FigTempMap}. The horizontal thermal structure presents very weak meridional and longitudinal contrasts, with a maximum temperature of 95.5$\pm0.5$K in the Southern Hemisphere and a minimum temperature of 83$\pm0.5$K northward of 50°N. Two studies have investigated the thermal disturbance induced at the tropopause level by the GWS using CIRS nadir spectra. Analyzing data obtained on November 2009 and August 2011, \citet{Achterberg2014} measured a temperature increase in the 25°N--40°N latitude band attributed to the GWS, but the temperature increase was limited to pressures larger than 400~hPa with not measurable modifications of the zonally averaged temperature at the tropopause level. In the longitude-pressure cross sections presented by \citet{Fletcher2012} for July 8th and July 26th, the tropopause temperature oscillated between 80K and 85K away from the beacon, a value very similar to our retrieved temperature, but presented a warm anomaly of about 10K just underneath the beacon at 100~hPa, with a longitudinal FWHM of 40°. We do not detect this warm 10-K anomaly below the stratospheric beacon using the TEXES dataset. In fact, we rather measure a temperature about 3K colder underneath the beacon than elsewhere in the same latitudinal band.

\subsection{Vertical thermal structure}

\begin{figure}
\subfigure{
	\includegraphics[trim=00mm 0mm 0mm 0mm, clip,width=0.45\linewidth]{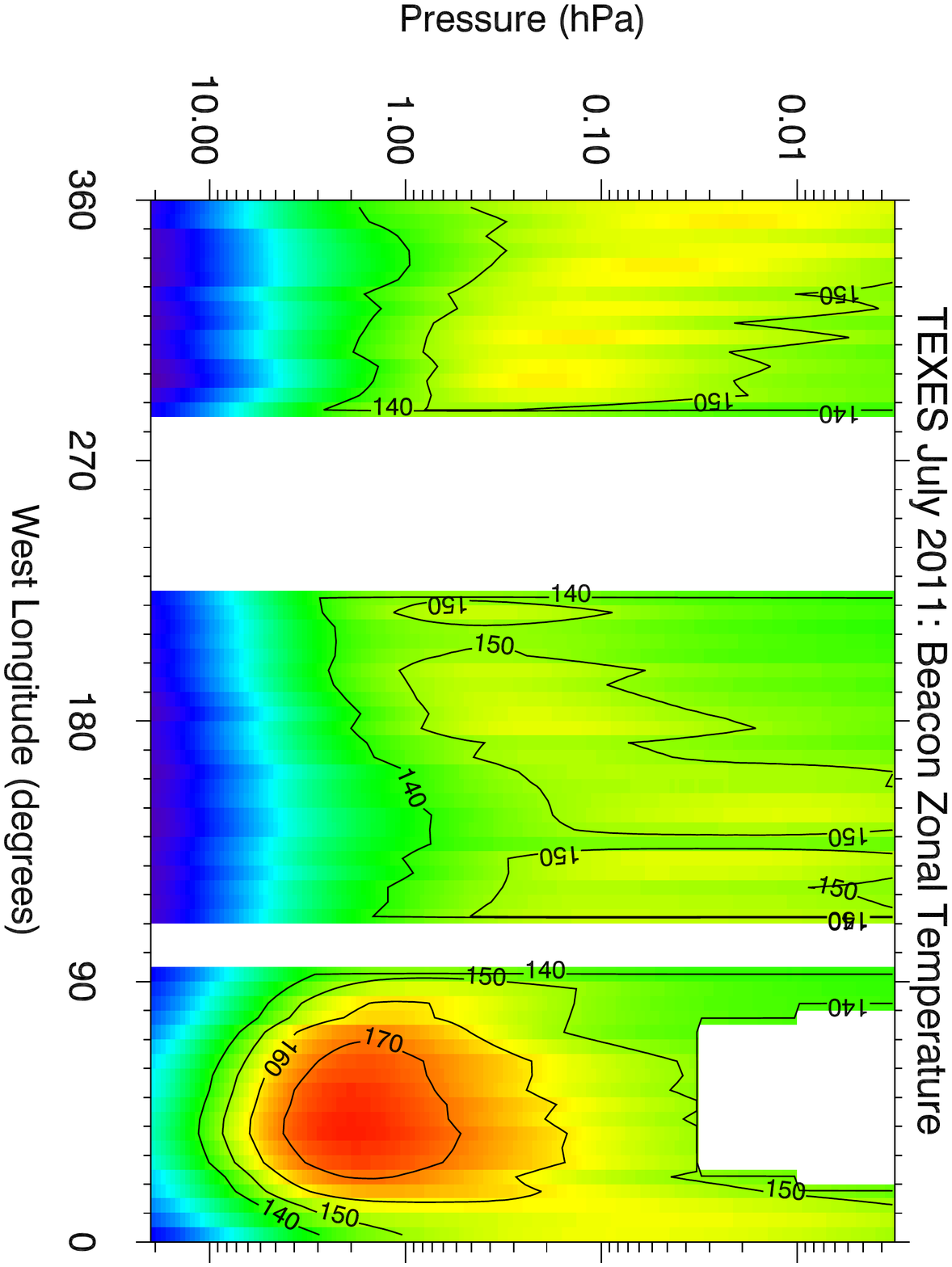}}
\subfigure{
	\includegraphics[trim=0mm 0mm 0mm 0mm, clip, width=0.45\linewidth]{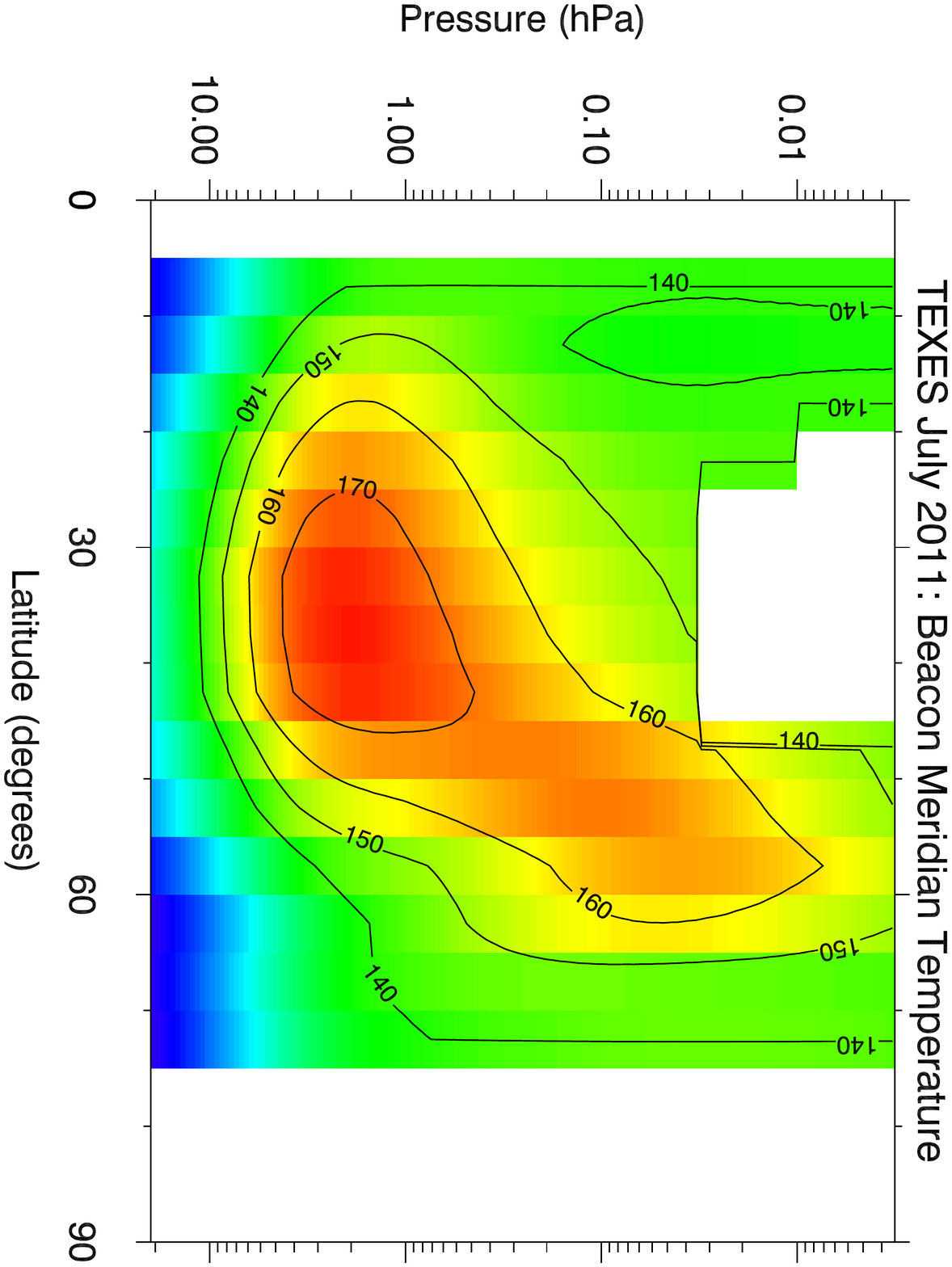}}
\subfigure{
	\includegraphics[trim=0mm 0mm 0mm 0mm, clip,width=0.45\linewidth]{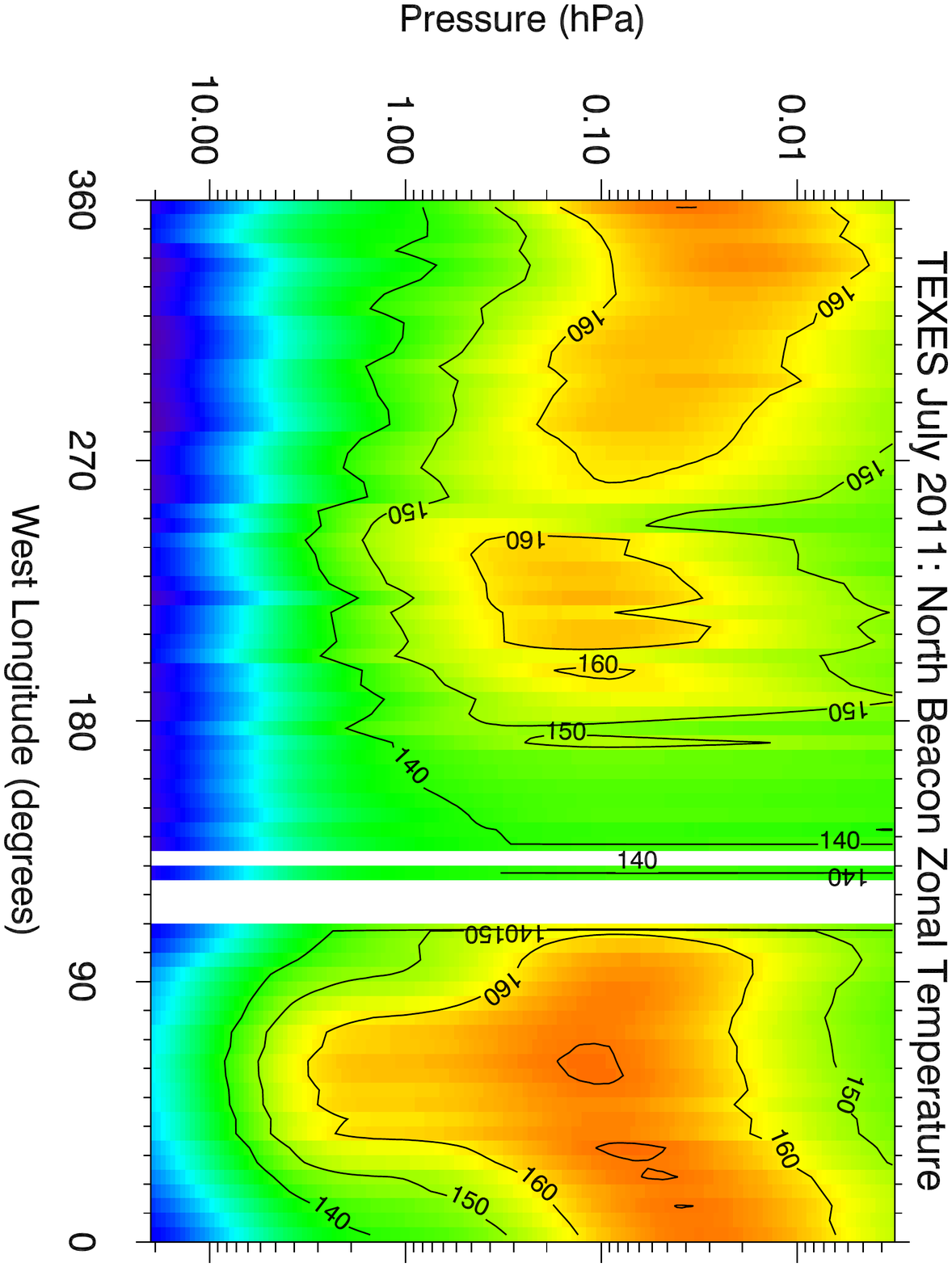}}
	\hspace{9mm}
\subfigure{
	\includegraphics[trim=0mm 0mm 0mm 0mm, clip, width=0.45\linewidth]{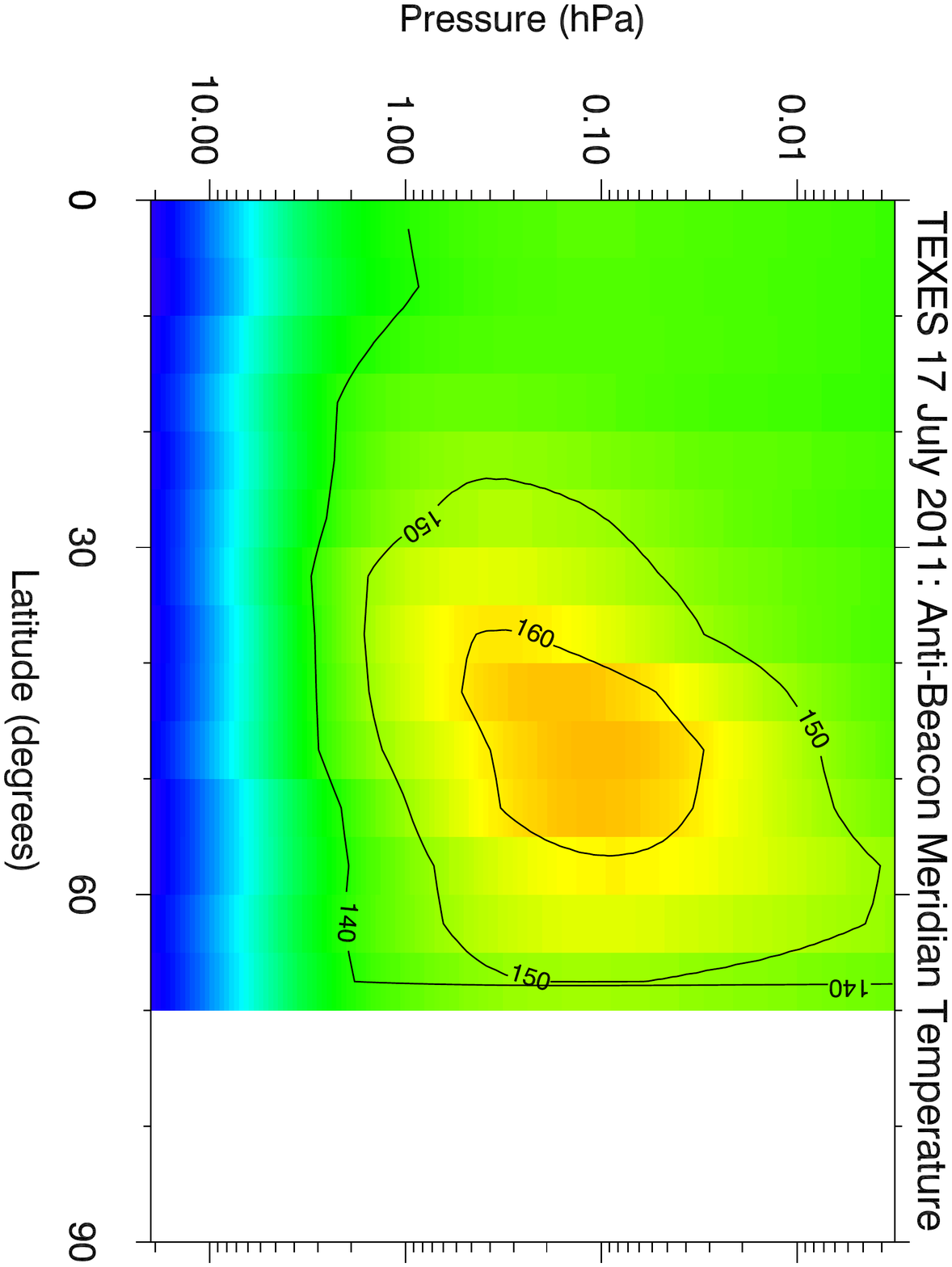}}
\caption{\label{FigCrossSections} Temperature cross sections retrieved for the 30°W--55°W zonal average on July 15th (upper left), for the 220°W-245°W  zonal average on July 17th (upper right), for the 30°N--45°N meridional average on July 15th \& 20th (lower left), and the  40°N--55°N meridional average on July 17th, 18th, and 19th (lower right).}
\end{figure}

To retrieve the stratospheric thermal structure at higher altitudes than the 0.1-hPa pressure level limit permitted by the signal-to-noise ratio of 5°$\times$5° binning in latitude and longitude, we coadded several CH$_4$ spectra to obtain four different zonal and latitudinal temperature cross-sections. As indicated by the dotted lines on the left panels of Fig.~\ref{FigTempMap}, two latitudinal cross-sections were obtained by averaging spectra in a 30°W--55°W stripe and a 220°W-245°W stripe respectively for July 15th and July 17th. These two cross-sections are displayed on the upper row of Fig.~\ref{FigCrossSections}. We also obtained two zonal cross-sections. The first one, displayed on the lower-left panel of Fig.~\ref{FigCrossSections}, runs from east to west at the latitude of the beacon, averaging spectra between 30°N and 45°N taken on July 15th \& 20th. The second zonal cross-section, displayed on the lower-right panel of Fig.~\ref{FigCrossSections}, runs through the warm upper stratosphere anomalies located to the north (40°N--55°N), using spectra taken on July 17th, 18th and 19th. The meridional swath and the zonal extent of the average for each date are represented by the dotted lines on Fig.~\ref{FigTempMap} and Fig.~\ref{FigTempMap19}. For the four cross-sections, we kept a 5° binning width along the cross-sections principal axes.

In the beacon, the meridional and zonal cross-sections displayed on the left column of Fig.~\ref{FigCrossSections} demonstrate that the maximum temperature anomaly is located at the 2-hPa pressure level uniformly over the beacon. The vertical extent of the beacon is rather uniform in longitude, but varies in latitude as it broadens by about one scale height between 25°N-30°N and 40°-45°N. Above the beacon (at pressures lower than 2~hPa), the temperature decreases rapidly with altitude, as it was already revealed by our qualitative analysis of the beacon's spectrum shown Fig.~\ref{Fig1245} (upper panel). A temperature of 160$\pm1$K is measured at 0.2~hPa and 35°N . Unfortunately, we are not in position to retrieve the temperature above the 0.1-hPa pressure level, as detailed in Sec.~\ref{SecRetrieval} and evident on the upper row of the Fig.~\ref{FigInfoCont1245}. Indeed, above this pressure level, the temperature profiles inverted from the two \textit{a priori} profiles start to deviate significantly form each other: the inversion using the cold \textit{a priori} profile relaxes towards 140K, while the inversion using the warm \textit{a priori} profile settles at a vertically uniform temperature of 160K. The fact that the temperature profile inverted starting from the warm \textit{a priori} does not relax to 180K at pressures lower than 0.1~hPa demonstrates that the atmosphere must be colder than 160K at these pressures.

Northward of the beacon (Fig.~\ref{FigCrossSections}, upper left panel), the pressure of the maximum temperature anomaly decreases from 2~hPa at 42.5°N down to 0.04~hPa at 57.5°N. In altitude, this corresponds to a steep slope of about four scale heights (i.e.\ 270~km) in 15° of latitude. This northward-upward tail of the beacon was already evident in CIRS nadir temperature maps obtained by \citet{Fletcher2012}, but these authors could not determine precisely its vertical structure because of the CIRS limited vertical sensitivity. They just noted that there were hints of warm stratospheric structures associated with the beacon at pressures lower than 0.1~hPa. Moreover, the temperatures we measure in this region, in excess of 160K, are much larger than the temperatures of 140K that were measured a few months before the GWS outbreak from CIRS limb spectra \citep{Sylvestre2015}. Therefore, this thermal perturbation north of the beacon is fully part of the GWS stratospheric aftermath.

\begin{figure}

\includegraphics[trim=0mm 0mm 0mm 0mm, clip, width=0.45\linewidth]{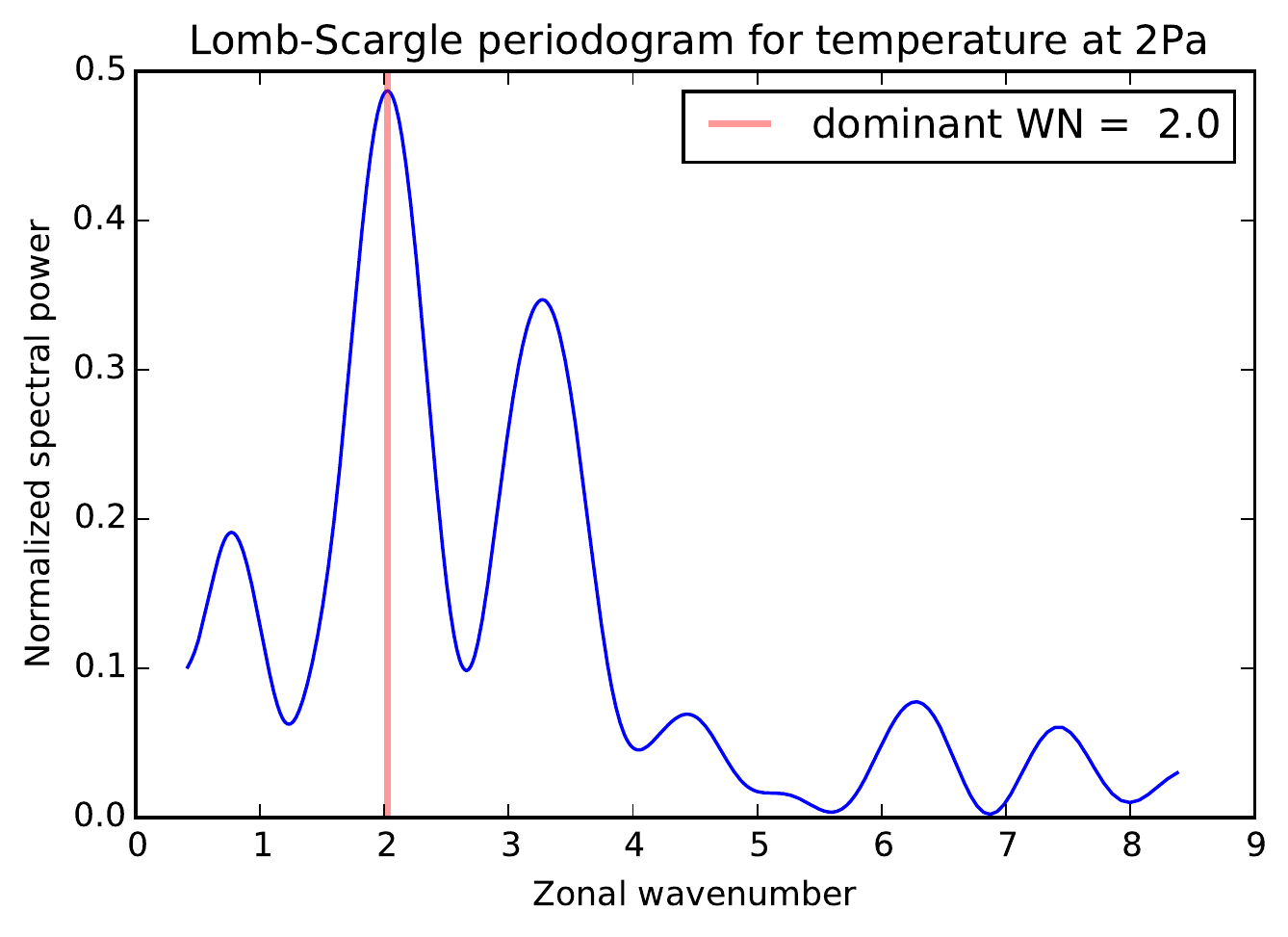}
\caption{\label{FigPeriodogram} Lomb-Scargle frequency analysis of the zonal temperature profile at 0.02~hPa in the 30°N--45°N meridional average (lower left panel of Fig.~\ref{FigCrossSections}).}
\end{figure}

To the east and west of the beacon (Fig.~\ref{FigCrossSections}, lower left panel), the measured temperature at the 2-hPa pressure level, about 140$\pm1$K, is similar to the temperatures observed before the GWS by CIRS nadir or limb spectra, but a thermal anomaly has developed at much lower pressures. It appears as a bubble of warm air undulating vertically with longitude.  Indeed, just to East of the beacon, at 357.5°W, the maximum temperature is reached at a pressure of 0.02~hPa, then slopes downwards away to reach the 0.2-hPa pressure level at 290°W, and rises again to pressures in range 0.01-0.05~hPa at 150°W. We have performed a Lomb-Scargle frequency analysis of the zonal temperature profile at 0.02~hPa to search for some periodicity in this structure \citep{Lomb1976,Scargle1982}. The periodogram, displayed Fig.~\ref{FigPeriodogram}, shows that the disturbance is dominated by a wavenumber-2 oscillation, but is not monochromatic as the wavenumber-3 also contributes to the signal. We must stress that our zonal cross-section was not obtained from a single observing snapshot, but rather by combining observations obtained on different days. Therefore, some structures may have moved with respect to each other, affecting our frequency analysis.

The temperatures measured in this warm bubble range between 150K and 160K. Similar results were obtained by \citet{Guerlet2013} from CIRS limb spectra taken in July and August 2011 at 40°N. At 60° East of the beacon center, these authors measured a maximum temperature of 170$\pm2$K at 0.05~hPa, while at 160° East of the beacon center, they measured a maximum of 160$\pm2$K located at 0.3~hPa. These high temperatures are 10K to 20K warmer than the temperatures measured from CIRS limb spectra at the same pressure levels before the storm \citep{Sylvestre2015}, and must also be seen as a consequence of the GWS.

The zonal temperature cross-section obtained in a stripe located to the North of the beacon is displayed in the lower right panel of Fig.~\ref{FigCrossSections}. In this panel, the only longitudinal range that appears not affected by the GWS is situated between 120°W and 180°W, west of the beacon. The rest of the stratosphere at this latitude has been affected, especially at low pressures. Indeed, the pressure where the temperature profile reaches its maximum is always lower than 0.2~hPa for all longitudes. At the beacon latitude, the maximum temperature occurs at a pressure of 0.1~hPa, rising to the 0.02-hPa pressure level at 340°W, and then decreasing to settle at 0.1-0.2~hPa between 250°W and 210°W. In this longitudinal range, the meridional cross-section displayed in the upper right panel of Fig.~\ref{FigCrossSections} shows that the pressure level of maximum temperature is not uniform with latitude, rising from 0.3-hPa at 42.5N to 0.1-hPa at 52.5°N.

 In summary, in the middle stratosphere, the temperature structure appears only altered by the beacon centered at the 2-hPa pressure level. In contrast, in the upper stratosphere, the thermal structure appears globally altered, although with a smaller amplitude than in the middle stratosphere, except directly to the west of the beacon. The pressure level of the maximum temperature perturbation oscillates zonally between 0.1--0.2~hPa above the beacon and 210°W-250°W and 0.02~hPa between 340°W and 360°W, and 130°W and 180°W. The pressure level of the maximum temperature perturbation also appears to rise northward at all longitudes.

\subsection{The CH$_4$ $2\nu_4-\nu_4$ hot lines}
\label{SecHotLines}

\begin{figure}
	\includegraphics[trim=0mm 0mm 0mm 0mm, clip, width=12cm]{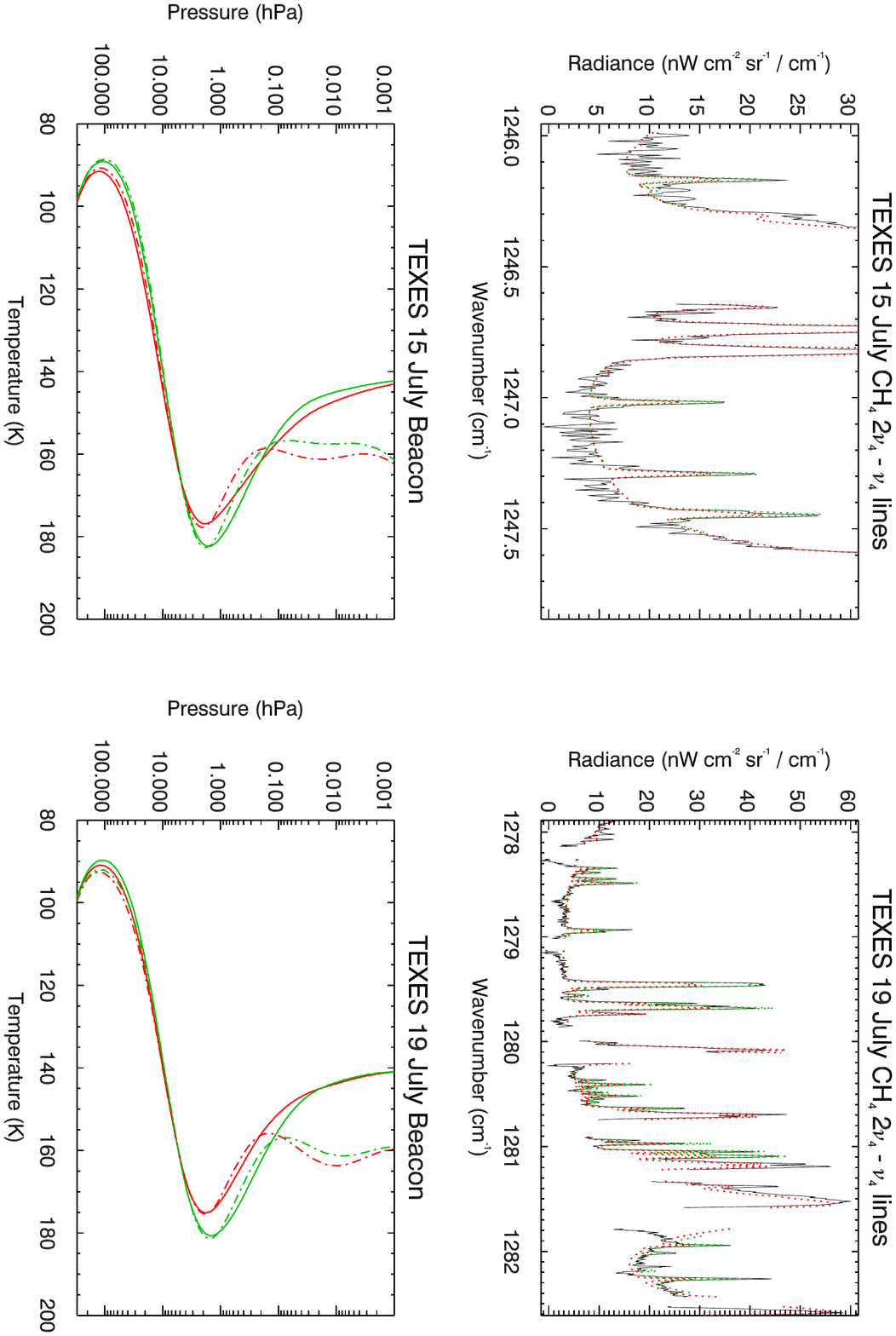}
	\caption{Upper panel: Comparison between the July 15th TEXES spectrum (black thick line) average over the beacon in the range 30°N--45°N and 30°W--55°W and two synthetic spectra: in red, the best-fit model to the full spectrum,  in green, the best-fit model the spectrum restricted to four lines belonging to the $2\nu_4-\nu_4$ hot band. Lowe panel: Vertical temperature profiles inverted from the full TEXES average spectrum (red lines) and the TEXES average spectrum restricted to the hot lines (green lines) for the cold \textit{a priori} (solid line) and the warm \textit{a priori} (dot-dashed line) profiles.}
	\label{FigHotLines}
\end{figure}

As presented in the upper panels of Fig.~\ref{Fig1245} and Fig.~\ref{Fig1280}, and highlighted by the zoom displayed in the upper row of Fig.~\ref{FigHotLines}, several CH$_4$ weak lines observed within the beacon are not properly reproduced by our synthetic spectra, and this for both the 1245-cm$^{-1}$ and 1280-cm$^{-1}$ settings. The synthetic radiance is systematically 30\% weaker than the observed radiance for each of these lines, which belong to the $2\nu_4-\nu_4$ hot band of methane, with upper rotational levels comprised between $J_{sup}=3$ and $J_{sup}=8$. Even if they are weak at room temperature, their intensities have been well constrained from laboratory measurements, better than 5\% according to the Table~5 of \citet{Ouardi1996}. This uncertainty, uncorrelated from line to line, is unable to account for the systematic difference between the synthetic and observed spectrum.

Methane fluorescence has been detected on Saturn \citep{Drossart1998}, and could contribute to the radiance in these lines. Indeed, at pressures lower than 0.05 Pa, CH$_4$ is excited to high-energy vibrational levels by absorption of solar IR photons, and relaxes to the ground-state more rapidly by emission of several photons than by collisions. However, the CH$_4$ $2\nu_4-\nu_4$ hot lines are not observed close to the equator, where this fluorescence is expected to be larger than at mid-latitudes. Thus, we do not think that fluorescence emission can explain the difference between our synthetic model and the observations.

Since these overtone lines have a high ground-state energy level, they may point towards the presence of higher temperatures in the core of the beacon, with a horizontal extent left unresolved by the seeing of the TEXES instrument. In order to assess this possibility, we performed a temperature inversion on the beacon average spectra obtained on July 15th and July 19th, but restricted to the methane lines belonging to the $2\nu_4-\nu_4$ hot band. For the \textit{a priori} temperature profiles, we used the profiles inverted from the respective full beacon average spectra for the cold and warm \textit{a priori}. The resulting temperature profiles are displayed in the lower row of Fig.~\ref{FigHotLines}. They all yield temperatures warmer by about 5K in the 1--2~hPa pressure range than the temperatures inverted from the full 1245-cm$^{-1}$ and 1280-cm$^{-1}$ spectra. In particular, the temperature measured using only the $2\nu_4-\nu_4$ hot lines within the 1245-cm$^{-1}$ setting points towards a maximum temperature of 185$\pm1$K at a pressure of 2~hPa, rather than the 180$\pm0.5$K retrieved at this level from the full spectrum. Such a temperature is still colder, but closer to the temperatures in the 190--200K range measured by CIRS \citep{Fletcher2012} in July--August 2011.

\subsection{TEXES spatial resolution: comparing the CIRS and TEXES datasets}
\label{SecCompCIRS}

To further investigate the differences in the temperatures inverted from the TEXES and CIRS datasets, we directly compared the two datasets. The TEXES dataset provides simultaneously a high spectral resolution and a broad spatial coverage, two assets that the CIRS dataset cannot match simultaneously. On its side, the CIRS dataset provides a much finer spatial resolution than TEXES, and is not affected by the terrestrial transmission. To address the differences between the two datasets in the spectral and spatial domains, we chose to compare the TEXES spectra obtained on July 15th using the 1245-cm$^{-1}$ setting with two different CIRS observations. The first observation, identified by 150SA\_COMPSIT001, was obtained on July 7th \&\ 8th at the highest CIRS spectral resolution, 0.5 cm$^{-1}$, and covered a full longitudinal band between 30°N and 40°N. The second observation, identified by 152SA\_FIRMAP001, was obtained later, on August 21st \&\ 22nd, at the lowest CIRS spectral resolution, 15.5~cm$^{-1}$, but it covered the entire Northern Hemisphere.

\begin{figure}
	\includegraphics[trim=20mm 85mm 10mm 80mm, clip, width=15cm]{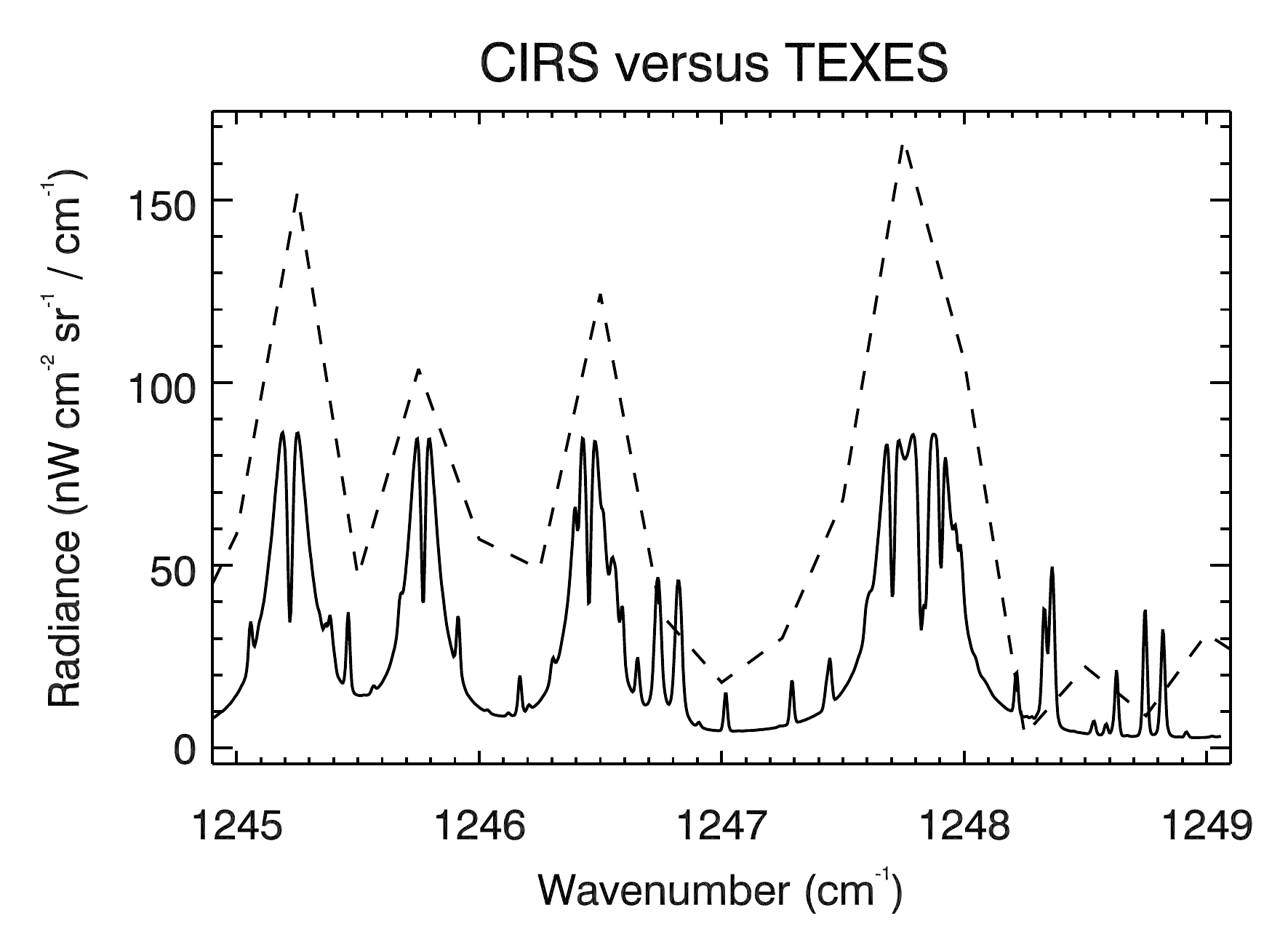}
	\caption{Comparison between the brightest CIRS spectra (dashed line) obtained over the beacon on July 7th-8th, at a spectral resolution of 0.5~cm$^{-1}$, and a synthetic spectra (solid line) adjusting the brightest TEXES spectra obtained over the beacon on July 15th.}
	\label{FigCIRSvsTEXES}
\end{figure}

Figure~\ref{FigCIRSvsTEXES} compares the brightest CIRS spectrum extracted from the sequence 150SA\_COMPSIT001 to the synthetic model best fitting the brightest TEXES spectra obtained on July 15th. This comparison circumvents the incomplete spectral coverage of TEXES due to opaque terrestrial atmospheric absorption. The strongest CH$_4$ lines at 1245.22, 1245.77, and 1246.45~cm$^{-1}$, and the manifold at 1247.8~cm$^{-1}$ are readily identified in both spectra. Strikingly, the CIRS spectrum obtained at a spectral resolution power of about 2,500 is still brighter than the TEXES spectrum obtained at a spectral resolution power of 75,000. Over the 1245--1249~cm$^{-1}$ spectral range, the CIRS mean radiance, 67.8~nW.cm$^{-1}$.sr$^{-1}$/cm$^{-1}$, is 2.4 times larger than the mean TEXES radiance of 28.0~nW.cm$^{-1}$.sr$^{-1}$/cm$^{-1}$. Such a large difference actually corresponds to the difference of the Planck function at 1247~cm$^{-1}$ for two black bodies at 180K, the warmest temperature by TEXES, and 200K, the warmest temperature measured by CIRS. It is however difficult to attribute such a large difference to the absolute calibration of both instruments, reliable within $\pm20\%$.

To investigate whether such a large factor could be explained by a difference in spatial resolution between the CIRS and TEXES datasets, we spatially convolved the CIRS sequence 152SA\_FIRMAP001 to mimic the spatial blurring affecting the TEXES dataset due to atmospheric transmission, and telescope and instrument diffraction. Since this CIRS sequence was obtained at a spectral resolution of 15.5~cm$^{-1}$ and a sampling step of 5~cm$^{-1}$,  it is irrelevant to directly compare the CIRS spectra and TEXES spectra. To perform the comparison, we mapped the mean CIRS radiance at 1245--1250~cm$^{-1}$ on Saturn's sphere. Then, we projected the Saturn's sphere onto the sky. In this process, we accounted for the westward drift of the beacon, 100° of longitude between July 15th and August 22nd, by projecting the CIRS data onto the sky with the same longitude offset between the beacon center and the central meridian as that of our TEXES observations. The CIRS radiance projected onto the sky was then convolved with a 2D Gaussian point-spread function. We varied the point spread function FWHM until the peak convolved CIRS radiance  was reduced by a factor of 2.4 compared with the peak unconvolved radiance. Such a reduction was obtained with a FWHM of 1.5 arcseconds.

As the result, the temperature we infer for discrete spatial features smaller than or comparable to 1.5\arcsec\ is lower than that inferred from CIRS spectra. This is relevant for the maximum temperature inferred within the beacon, $180\pm1$K compared to  $200\pm1$K find by \citet{Fletcher2012}, but also to the warm anomalies in the upper stratosphere, where we infer temperature up to 160K while \citet{Guerlet2013} found a maximum temperature of $170\pm2$K. This conclusion is also consistent with our inversion of the CH$_4$ $2\nu_4-\nu_4$ hot lines (not resolved at CIRS 0.5-cm$^{-1}$ spectral resolution) from which we retrieved a $185\pm1$K maximum temperature. The $2\nu_4-\nu_4$ hot lines have a much sharper dependence on temperature than the $\nu_4$ fundamental lines. Hence, relatively to the background radiance, the radiance of the hot lines in the core of the beacon must be stronger than that of $\nu_4$ fundamental lines. Then the convolution of the hot lines by the IRTF seeing led to a retrieved temperature warmer than that of the $\nu_4$ lines.

We also investigated whether the difference in spatial resolution could explain the difference in the inferred temperature at the tropopause level between CIRS and TEXES. At 610~cm$^{-1}$, CIRS measured a radiance three times brighter at beacon's latitude and longitude on August 21st \&\ 22nd (sequence 152SA\_FIRMAP001). This anomaly had a FWHM of 40° in longitude, hence about two times smaller than the stratospheric anomaly which had a 80° of longitude FWHM. In contrast, TEXES observed a smaller contrast, a factor of about 1.5, between the measured radiances at beacon's position and in the surroundings. The CIRS and TEXES radiance contrasts can be reconciled by convolving the CIRS dataset with a seeing of $\sim2$ arcseconds. This larger seeing for the H$_2$ setting than for the CH$_4$ setting is expected as we used a wider slit for the H$_2$ setting than for the CH$_4$ setting. Since the thermal anomaly at the tropopause had a larger FWHM than the thermal anomaly in the stratosphere, the effective contrast in the TEXES dataset of the former was more reduced than that of the latter. In the inversion process, the warm stratosphere needed to fit the strong H$_2$ quadrupolar line within the beacon raised the continuum level more than was observed by TEXES, a defect that the algorithm compensated by slightly decreasing the tropopause temperature.

We thus conclude that the CIRS and TEXES datasets are in agreement with each other assuming seeings of 1.5 arcseconds for the CH$4$ setting and 2 arcseconds for the H$_2$ setting at the time of our TEXES observations on July 15th. Such values seem sensible for late afternoon observations at the Mauna Kea. Nevertheless, as the exact seeing was not monitored independently at the IRTF, we cannot exclude that offsets in radiometric calibration could also be contributing.

\section{Discussion}
\label{SecDiscussion}

In this section, we  discuss how the TEXES dataset we have presented may help to decipher by which mechanisms the 2010 Great White Storm disturbed Saturn's stratospheric thermal structure. The thermal structure we have measured adds some new information to the previous knowledge on the stratospheric state in mid July 2011. Our study definitely demonstrates that the thermal perturbation can be divided into two independent disturbances, lying at different pressure levels. The main disturbance was already identified by CIRS and VISIR/VLT observations \citep{Fletcher2011,Fletcher2012} as the B0 beacon. It is a warm vortex located in the mid-stratosphere. The second anomaly, is weaker, horizontally more extended than the beacon, and located in the upper stratosphere ($p<0.2$~hPa). This anomaly was also identified in the CIRS dataset, but incorrectly located at the 0.5-hPa pressure level.

\subsection{Beacon}

The first characteristic of the B0 beacon that remains to be explained is its vertical structure. Our TEXES dataset confirms the CIRS and VISIR/VLT finding that the beacon thermal anomaly was centered at 2~hPa in mid-July 2011. This center pressure had changed during the evolution of the beacon, as the two early disturbances B1 and B2 were initially centered at 0.5~hPa, before their merge resulted in a downward shift to the 2-hPa pressure level \citep{Fletcher2012}. In July 2011, the high spectral resolution power of TEXES allows us to conclude that, above this pressure level, the temperature was decreasing with altitude at least up to the 0.2-hPa pressure level. Unfortunately, we cannot constrain firmly the temperature profile at pressures lower than 0.2~hPa. We can only state that it was cooler than 160K.

A second distinctive feature of the beacon is its horizontal confinement. This can be readily seen by inspecting profiles of the potential temperature $\theta=T(p_0/p)^{\kappa}$, where $T$ is the temperature at the pressure $p$, $p_0$ the reference pressure ($p_0=1000$~hPa), and $\kappa$ is defined from the specific heat at constant pressure $c_p$ and the universal gas constant $R$: $\kappa = R/c_p$.  In Fig.~\ref{FigTempPot}, we compare potential temperature profiles measured from the TEXES dataset to that obtained from the CIRS limb dataset acquired before the storm outbreak, hence pertaining to the quiescent stratosphere \citep{Sylvestre2015}. 
In the 30°N--45°N latitudinal band, the potential temperature outside of the beacon remained largely unchanged compared to pre-storm conditions, while within the beacon, the potential temperature corresponded to that before the onset of the GWS at 1~hPa, i.e.\ 0.7 scale higher. This demonstrates that, in the mid-stratosphere, the diabatic heating or the vertical advection of potential temperature generated by the GWS only took place within the beacon. The beacon can be seen as a thermal anomaly sitting in a quiescent, unaltered background. The potential temperature meridional profile shows that the beacon anomaly is centered on the northern edge of a region of low baroclinicity prior to the storm outburst.

The potential temperature inferred within the beacon demonstrates that this warm anomaly cannot be due to an overshooting of the deep tropospheric convection through the tropopause. If it were the case, the potential temperature should be conserved throughout the stratosphere and equal to the temperature at $p_0=1000$~hPa (1 bar), about 160K. This conclusion is also consistent with the fact that the tropospheric cloud layer and the tropospheric temperatures were not affected above the 300-hPa level \citep{SanzRequena2012,Achterberg2014}. As already proposed by \citet{Fletcher2012}, the beacon should rather be a stratospheric response to the dynamical forcing that the convective storm head imposed on the stably stratified layers of the upper troposphere and stratosphere. \citet{Fletcher2012} especially studied how topographically forced Rossby waves may have formed the beacon. Here, we investigate whether transport of energy and momentum by gravity waves could have produced the observed heating. Indeed, based on an analogy with well-known phenomena in the Earth's stratosphere \citep{Alexander1997}, the convective outburst causing the Great White Spots should have emitted a large spectrum of intense gravity waves when impinging the stable layers in the tropopause region.

We first need to consider in which part of the stratosphere a significant gravity-wave activity could have been triggered by the GWS.  Gravity-wave dispersion is mainly determined by two atmospheric variables, the  mean zonal flow $\langle u \rangle$, and the atmospheric static stability represented by the Brunt-Väisälä frequency $N$. Their effects can be seen by inspecting the dispersion relation of a gravity wave
\begin{equation}
k_z^2 = k_x^2\left[ \frac{N^2}{(\omega-\langle u \rangle k_x)^2} - 1 \right]
\end{equation}
where $k_x$ and $k_z$ are the zonal and vertical wavenumber of the wave, $\omega$ its frequency. This equation holds when compressibility and Coriolis effects are negligible, and when the wave vertical wavelength is small compared to the local atmospheric scale height ($Hk_z\gg 1$). In this equation, if $N^2$ decreases, or if the term $(\omega-\langle u \rangle k_x)^2$ increases due to a change in the mean zonal flow, $k_z^2$ may become negative and the wave evanescent.


The mean zonal flow has never been measured directly in Saturn's stratosphere, but it has been inferred indirectly from the temperature structure, assuming gradient-wind equilibrium: the vertical shear $\partial \langle u \rangle /\partial z$ is proportional to the temperature meridional gradient. Using the CIRS limb spectra, \citet{Sylvestre2015} found, just before the onset of the GWS, a weak temperature meridional gradient in the Northern Hemisphere at all pressures in the range 0.01--1~hPa. This situation, suggesting weak vertical changes of zonal wind speed, was propitious to upward propagation of gravity waves. \citet{Fletcher2011} also derived the stratospheric gradient wind before the GWS using the CIRS nadir spectra, which provide a finer meridional sampling  than the CIRS limb spectra. As shown in their Fig.~S3, the early beacons B1 and B2 started at latitudes, respectively 40°N and 25°N (45°N and 27°N in the planetographic coordinates used by \citet{Fletcher2011}), where the zonal jets derived from gradient wind equilibrium were, prior to the storm, nearly constant with altitude. In contrast, no temperature anomaly developed in the 25°N-40°N latitude band, where the zonal gradient winds were increasing with altitude. Hence, the inspection of Saturn's thermal structure suggests that the beacons developed at latitudes where upward propagation of gravity waves was favored by a weak vertical shear of the mean zonal flow $\langle u \rangle$. At other latitudes, gravity waves may have been reflected downwards or refracted by atmospheric baroclinicity \citep{Nappo2002}. 

The static stability $N$ is directly related to the potential temperature vertical gradient through the equation 
\begin{equation}
N=\sqrt{ \dfrac{g}{\theta} \dfrac{\partial\theta}{\partial z} }
\end{equation}
where $g$ is the acceleration of gravity.
The CIRS limb observations at northern mid-latitude have shown that, at about 1~hPa, the temperature vertical gradient is changing from rapidly increasing with altitude below this pressure level to moderately increasing or constant with altitude above this pressure level \citep{Guerlet2009,Sylvestre2015}. The same change in vertical gradient was observed in the temperature profiles derived from the Voyager radio occultations \citep{Lindal1985}, from which the \textit{a priori} profiles displayed in Fig.~\ref{FigInfoCont1245} are derived. This change in vertical temperature gradient, inducing a decrease of the static stability $N$, could make gravity waves with  horizontal wavelength smaller than a critical wavelength evanescent, and could limit their activity to pressures higher than about 1~hPa. Upward-propagating gravity waves were reflected downward, or forced to propagate horizontally, when they reached this pressure. Hence, our qualitative analysis leads us to conclude that a significant gravity-wave activity only took place in the lower stratosphere ($p\gtrsim 1$~hPa), in two narrow latitude bands centered at 25°N and 40°N. 

\citet{Harrington2010} reported the detection of temperature fluctuations in their temperature vertical profile inverted between 0.1 and 6~Pa from the occultation of the star by GSC 0622-00345 by Saturn. They interpreted these temperatures as the signatures of gravity waves, although they could not exclude sound waves or planetary waves as the cause of the temperature fluctuations. They measured the waves vertical wavelengths and, using the method of \cite{Raynaud2004}, derived a model-dependent horizontal wavelength. However, they did not determine the wave phase speed relative to the zonal mean flow, $(c_x-\langle u \rangle)$ a critical parameter for our study. In the absence of other report on gravity waves in Saturn's atmosphere, we adopt a mean value of $(c_x-\langle u \rangle)$ of 300 m/s derived for gravity waves observed in Jupiter's atmosphere. For the Brunt-Väisälä frequency of 0.25 s$^{-1}$ at 1~hPa, this value $(c_x-\langle u \rangle)$ makes gravity waves with horizontal wavelength smaller than 7.5~km evanescent in Saturn's atmosphere. Supporting this rough estimate, it is interesting to note that \citet{Harrington2010} did not detect waves with horizontal wavelength smaller than 10~km.

We now consider how the gravity-wave activity may have warmed the stratosphere. Gravity waves deposit their energy and momentum when their amplitudes become larger than the potential temperature variation of the background atmosphere over a wavelength.  Neglecting compressible and Coriolis effects, gravity-wave breaking can be formulated through the saturation index \citep{Hauchecorne1987,Spiga2012}
\begin{equation}
\mathcal{S} = \sqrt{\frac {\alpha N}{\langle \rho \rangle |\langle u \rangle-c|^3} } \qquad \text{with} \qquad \alpha = \frac{F_0}{k_z} 
\label{EqSaturation}
\end{equation}
where $\langle \rho \rangle$ is the background density, $c$ the gravity-wave phase speed, and $F_0$ the gravity-wave vertical momentum flux (conserved for non-dissipating gravity waves). If $\mathcal{S}$ approaches 1, the more likely the gravity wave is to saturate and break. In contrast, if $\mathcal{S}$ remains significantly smaller than 1, the gravity wave can propagate upwards.

Equation~\ref{EqSaturation} shows that gravity-wave breaking is favored by large static stability $N$, and by small mean atmospheric density $\langle \rho \rangle$. It is therefore reasonable to assume that gravity waves predominantly broke around the 1-hPa pressure level, at the top of the highly stable region where significant gravity-wave activity was possible for Saturn's background atmosphere. This 1-hPa pressure level lies close to the critical pressure level where the maximum stratospheric heating took place, in the pressure range 0.5--2~hPa.

The initial heating may then have triggered a positive feedback mechanism.  This initial heating changed the temperature vertical gradient, raising the static stability below the heating altitude. Moreover, the gravity waves depositing their momentum may have also locally decelerated the mean zonal flow, so that $\langle u \rangle - c$ has decreased. The combination of the two effects may then have resulted, according to Eq.~\ref{EqSaturation}, in a strong increase in gravity-wave saturation, and in a preferential breaking of the waves below the initial heating. This positivive feedback mechanism may have generated and strengthened the original beacons. It may also explain why the heating only took place within the beacons and left the rest of the longitudinal band hardly affected (Fig.~\ref{FigTempPot}). This positive feedback mechanism has been identified in the terrestrial mesosphere as responsible for the presence of strong inversion layers during several consecutive days \citep{Hauchecorne1987,Vanz1989}. Moreover, gravity waves tend to break in the region of maximum vertical temperature gradient, i.e.\ below the level of maximum temperature. As a result, gravity waves would not have been able to break and warm the layers situated above the beacon maximum temperature, consistently with the TEXES observations showing a decrease of the temperature between 2~hPa and 0.2~hPa. Instead, gravity waves induced a downward shift of the maximum temperature. This effect is extensively documented in the terrestrial atmosphere, where it drives the downward propagation of the stratospheric quasi-biennal oscillation \citep{Baldwin2001}. It may also have driven the downward shift from 0.5~hPa to 2~hPa of the beacon maximum temperature. Yet, we stress that the beacon downward shift seems to have occurred suddenly, when the beacons B1 and B2 merged, while our proposed mechanism would rather gives rise to a slower and steady shift.

In this paper, we will not go further than a qualitative discussion. Our scenario can only be validated by a Global Climate Model (GCM). In particular, our analysis relies on the gradient wind equation that may not hold in the presence of a strong wave activity, and, for this reason, it must be taken with caution, and should be validated by a numerical modeling. \citet{Fletcher2012} tried to use the EPIC atmospheric model to study gravity-wave propagation, but were limited by their coarse vertical resolution. Indeed, modelling gravity-wave propagation and breaking requires numerical models with much finer vertical resolution than currently carried out \citep{Friedson2012,Guerlet2014}. A high-resolution GCM will also validate whether the gravity-wave spectrum and flux emitted can transport as much energy and momentum as required to explain the stratospheric heating observed by TEXES and CIRS.

\subsection{The upper stratosphere}

\begin{figure}
\subfigure{
	\includegraphics[trim=0mm 0mm 0mm 0mm, clip,width=0.40\linewidth]{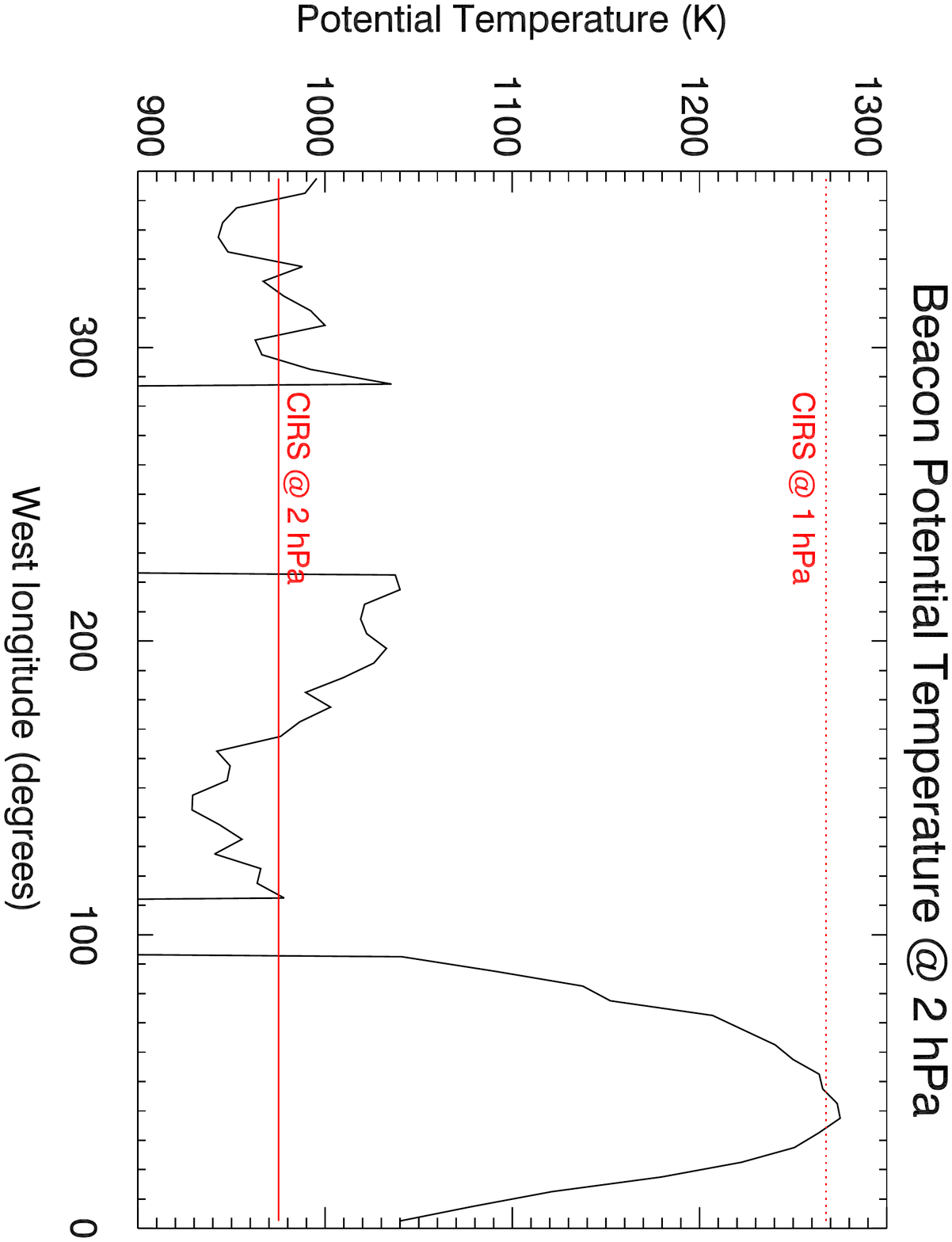}}
\subfigure{
	\includegraphics[trim=0mm 0mm 0mm 0mm, clip=true, width=0.40\linewidth]{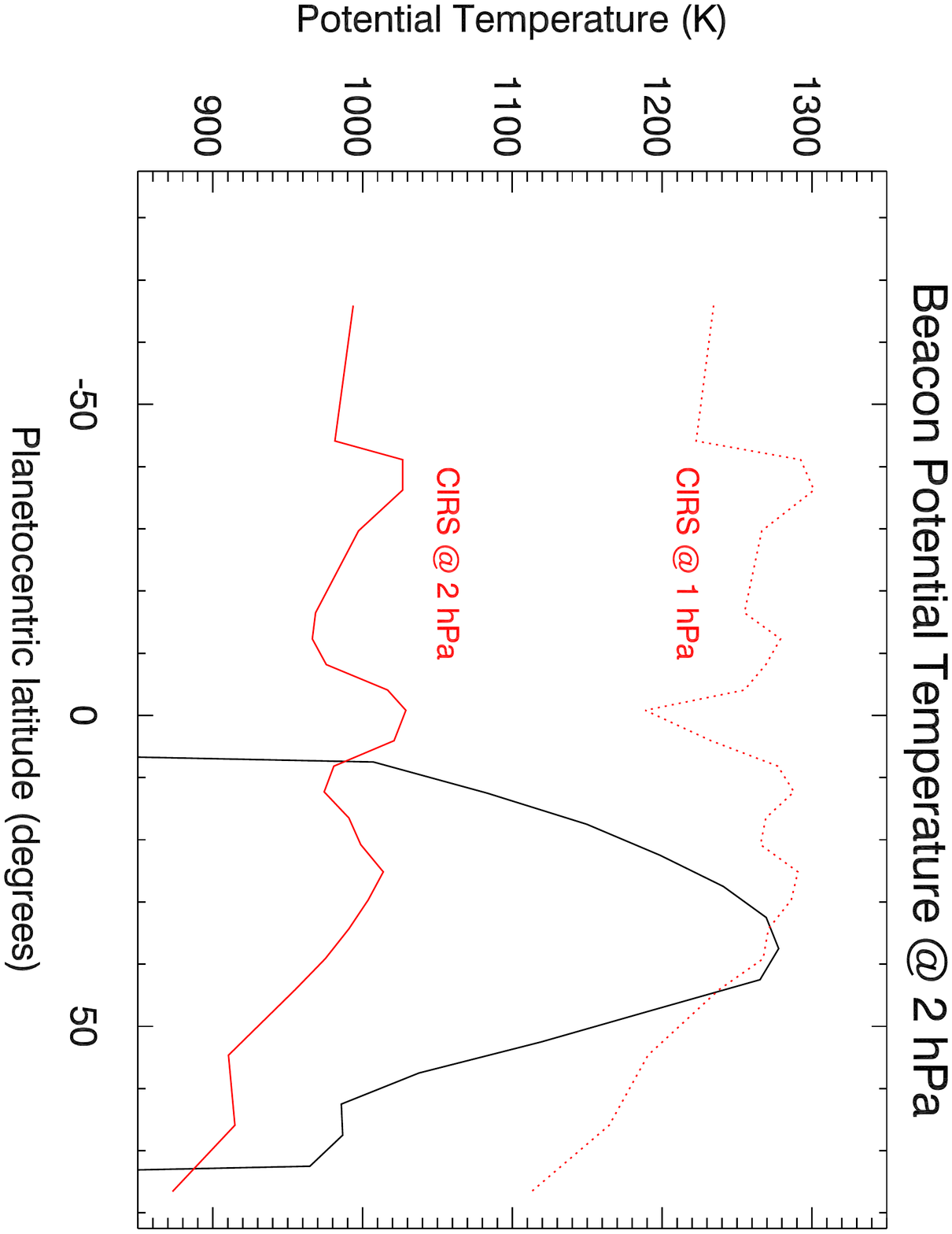}}
\subfigure{
	\includegraphics[trim=0mm 0mm 0mm 0mm, clip,width=0.40\linewidth]{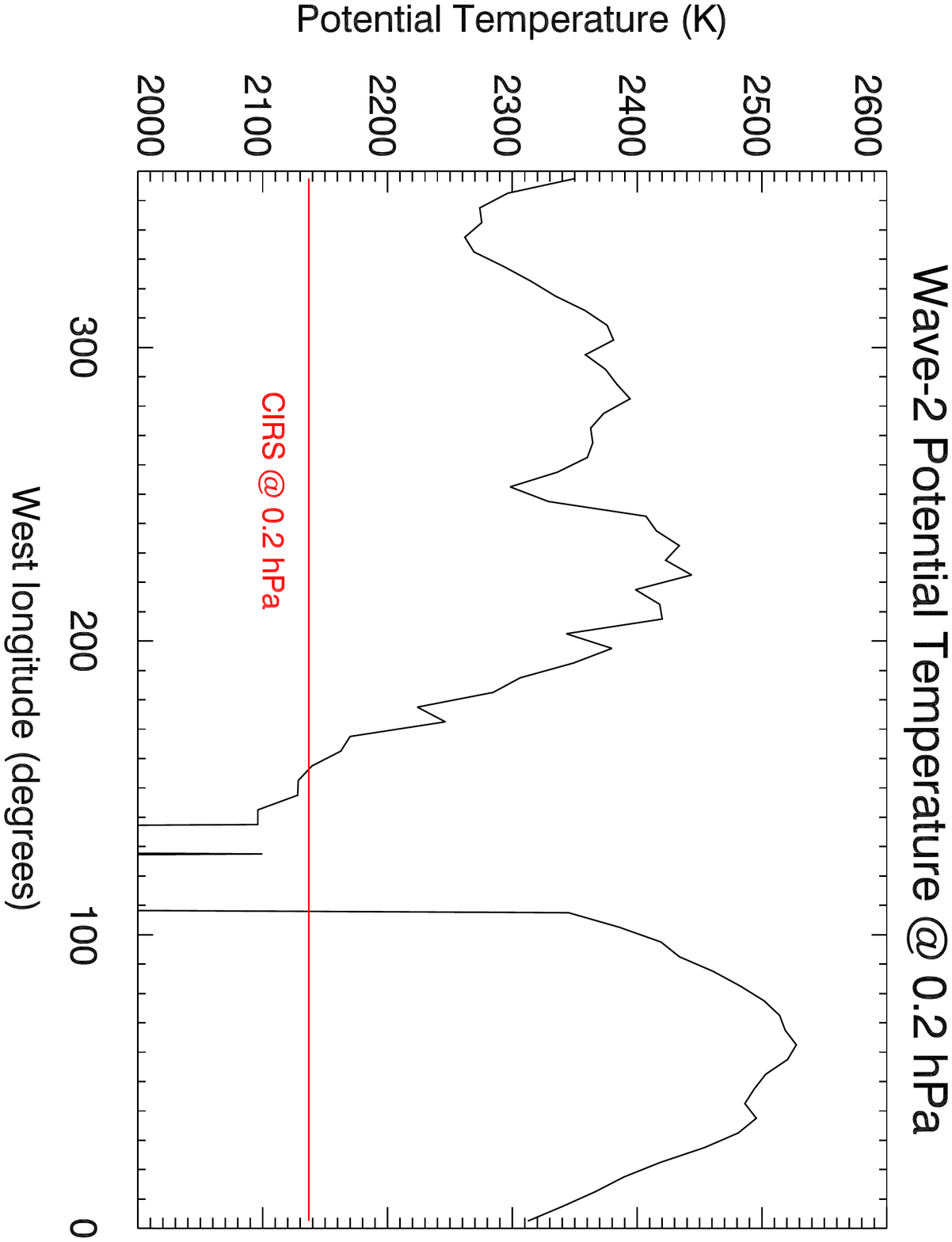}}
	\hspace{22mm}
\subfigure{
	\includegraphics[trim=0mm 0mm 0mm 0mm, clip, width=0.40\linewidth]{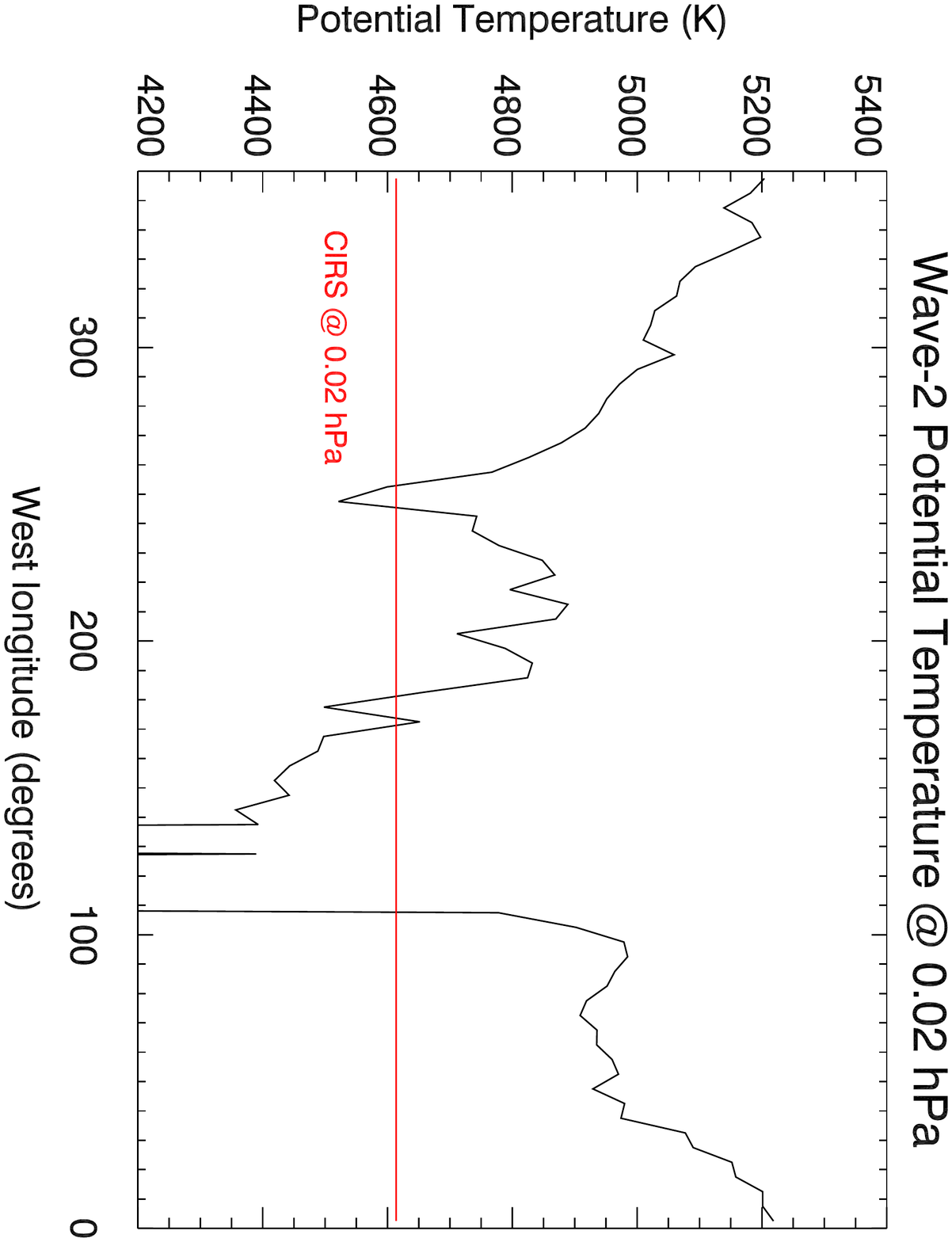}}
\caption{\label{FigTempPot} Zonal and meridional profiles of potential temperature calculated for the thermal structure measured from TEXES (black lines) and from CIRS limb observations (red lines). The upper left panel displays the meridional profile across the beacon corresponding to the temperature cross sections retrieved for the 30°W--55°W zonal average (Fig.~\ref{FigCrossSections} upper left) at 2~hPa. The lower left panel displays the zonal profile across the beacon corresponding to the temperature cross sections retrieved for the 30°W--55°W zonal average (Fig.~\ref{FigCrossSections} lower left) at 2~hPa. The right panels display the zonal profile corresponding to the temperature cross sections retrieved for the 40°N--55°N  average (Fig.~\ref{FigCrossSections} lower left) at 0.2~hPa (upper right) and 0.02~hPa (lower right).}
\end{figure}

In Saturn's upper stratosphere (pressure range~$0.1-0.01$~hPa), TEXES observations have shown that the pattern drawn by warmer versus colder areas
in Fig.~\ref{FigTempMap} and~\ref{FigCrossSections} is dominated by
a wavenumber-2 longitudinal planetary wave (Fig.~\ref{FigPeriodogram}). This wavenumber-2 planetary disturbance is not
observed in the pre-beacon Saturn's stratosphere. Some weak longitudinal variations have been observed, but a pressures around 1~hPa \citep{Orton2013}. The zonal profiles of the potential temperature at 0.2~hPa and 0.02~hPa derived from the thermal structure presented in the lower right panel of Fig.~\ref{FigCrossSections} are compared to the potential temperature retrieved using the CIRS limb spectra on the right panels of Fig.~\ref{FigTempPot}. They show that the TEXES oscillating potential temperature profiles have two minima slightly colder than CIRS pre-beacon observations, and two maxima warmer than the pre-beacon situation. 
Several CIRS nadir spectra and VISIR/VLT observations had identified this warm tail as apparently originating from the beacon. \citet{Fletcher2012} correctly stated that it was located at pressures lower than 0.5~hPa, but could not retrieve its correct altitude as they were limited in vertical sensitivity by their modest spectral resolution. If their observations did not regularly  sample this high altitude disturbance, they were nevertheless able to spot some temporal changes in its disturbance.

One possibility to explain this pattern is that the warm beacon area
exerts in Saturn's atmosphere a similar
forcing as the solar heating does in the
strongly radiatively-controlled 
Martian atmosphere, giving rise 
to thermal tides \citep{Wilson96}.
The longitudinal wave witnessed by TEXES
in Saturn's stratosphere would thus
be a thermal tide signal with 
a dominant semi-diurnal mode.
This possibility opens many perspectives
for the study of Saturn's atmosphere,
since its long radiative timescale
prevents it from being susceptible to thermal tides
forced by the diurnal cycle of solar heating pattern.
Another possibility is that this wavenumber-2 mode
arose because the anticyclonic vortex associated with the beacon caused a modification of
planetary wave activity, similar to the one
observed on Earth following sudden stratospheric
warmings \citep{Matsuno1971,Hoffmann2007}. 
Interestingly, terrestrial sudden stratospheric
warming causes zonal wind changes, with
possible weakening of the eastward jets
caused by westward acceleration 
-- this kind of change is also observed at altitudes
1~hPa and latitudes 40° 
on Saturn \citep[Fig~2C][]{Fletcher2011}. \citet{Andrews1987} also showed theoretically that an anticyclonic vortex and a warm surface can cause the same atmospheric response, both behaving like thermal tides.

Still, as in the case of the putative role of gravity waves in the generation of the beacon, we stress that our interpretation must be validated by a GCM. In fact, this study of thermal tides induced by the beacon could be performed more easily than that of the gravity waves as it requires a coarser spatial resolution than modeling the interactions of gravity waves with the mean zonal flow.

\section{Conclusions}

The high spectral resolving power of TEXES has allowed us to retrieve the entire vertical thermal structure (10--0.001~hPa) of the stratospheric disturbance generated by the 2010 GWS, as of mid-July 2011. The main disturbance, the beacon, appears as a warm airmass centered at 2~hPa. Our measured vertical structure, latitudinal and longitudinal extent, are consistent with the measurements performed by \citet{Fletcher2012}. Our retrieved maximum temperature $180\pm1$K is colder than that retrieved from the CIRS spectra ($200\pm1$K), but this difference can be accounted for by a seeing of 1.5\arcsec, typical for IRTF observations performed during the afternoon or at the beginning of the night. Our observations also suggest that the beacon longitudinal drift rate changed abruptly between July 15th and July 20th from (1.6°$\pm$0.2°)/day to (2.7°$\pm$0.04°)/day. At the beacon central latitude in July 2011, outside of the beacon itself, 2-hPa temperatures were roughly consistent with pre-beacon conditions. However, that is not the case for lower pressures such as the 0.2-hPa level, where thermal anomalies extended over a much larger longitude region. Northward of the beacon, the pressure level of the maximum temperature anomaly rises up to 0.04-hPa at 52.5°N. Zonally, the upper-stratosphere thermal anomaly appears as a dominant wavenumber-2 temperature perturbation affecting the entire longitude circle. The pressure of the maximum perturbation undulates between 0.2 and 0.02~hPa.

We propose a qualitative explanation for the generation of the beacon and of the upper stratosphere perturbation. We posit that gravity waves emitted by the tropospheric convective storm were able to propagate in a small part of Saturn's stratosphere, vertically limited by a change in static stability occurring around 1~hPa on the quiescent temperature vertical profile, and horizontally limited by the baroclinicity of the atmosphere. The breaking upward propagating gravity waves was favored by the low densities at the top of the stable region, at about 1~hPa, inducing an early warming. The early perturbation on the temperature and zonal wind then triggered a positive feedback that forced the gravity waves to break preferentially within the thermal anomaly, raising the temperature as long as the GWS was active, and the original beacons were collocated in longitude above tropospheric features that could have been the source of these gravity waves. The warm beacon then generated thermal tides this upper stratosphere perturbation, where the semi-diurnal mode dominated the diurnal mode, or simply planetary wave activity in the upper stratosphere.

This scenario remains qualitative and requires further testing with numerical models to check (i) if gravity waves were actually able to propagate in the stratosphere for the background atmosphere observed prior to the GWS, (ii) to check if they could have been damped at the observed altitude, and (iii) if the convective storm was able to generate the flux and spectrum of gravity waves required to reproduce the heating observed. Besides heating, breaking gravity waves would have other effects on the atmosphere, such as generating an intense turbulence that should have vertically mixed chemical species. This turbulence may be the explanation for the increase in hydrocarbons observed within the beacon by \citet{Fletcher2012,Hesman2012}, as suggested by \citet{Cavalie2015}. Large-scale downwelling winds, part of a residual circulation triggered by the beacon, may also have played a role \citep{Cavalie2015,Moses2015}. In the future, we will use TEXES observations of stratospheric hydrocarbons to monitor their enhancement within and outside of the beacon. Finally, our scenario can also be tested in the event of a future storm eruption. If the next GWS occurs at latitude where the vertical shear of the mean zonal flow hampers the upward propagation of gravity waves, this GWS should not affect the stratosphere in the same way as the 2010 GWS.

\section{acknowledgement}

T.\ Fouchet acknowledges support from the INSU/CNRS Programme national de planétologie and from the Institut universitaire de France.

T.\ Greathouse acknowledges support from NASA PAST grant NNX08AW33G used to retrieve the data and support for paper preparation from NASA PAST grant NNX14AG35G. 

G.\ Orton acknowledges support to the Jet Propulsion Laboratory, California Institute of Technology, from the National Aeronautics and Space Administration.

The authors thank Wesley Irons for participating to the observation campaign. The authors thank Emmanuel Lellouch for fruitful comments to the manuscript.

\section{References}
\bibliographystyle{elsarticle-harv}
\bibliography{/Users/tfouchet/Documents/Document_pc/latex/Articles/bibliographie}

\end{document}